\documentclass{article}

\PassOptionsToPackage{numbers, compress}{natbib}

\usepackage[nonatbib,preprint]{neurips_2023}
\usepackage{authblk}

\usepackage[normalem]{ulem}

\usepackage{setspace}
\usepackage{comment}

\newcommand{\bs}{\boldsymbol}
\def\RR{ \mathbb R}
\newcommand{\refeqp}[1]{Equation (\ref{#1})}

\newcommand{\ee}{\end{equation}}
\newcommand{\be}{\begin{equation}}
\newcommand{\ec}{\end{center}}
\newcommand{\bc}{\begin{center}}
\newcommand{\eea}{\end{eqnarray}}
\newcommand{\bea}{\begin{eqnarray}}
\newcommand{\bd}{\begin{description}}
\newcommand{\ed}{\end{description}}
\newcommand{\bi}{\begin{itemize}}
\newcommand{\ei}{\end{itemize}}
\newcommand{\pa}{\partial}

\newcommand{\bx}{\bs{x}}

\newcommand{\bz}{\bs{z}}

\newcommand{\bt}{\bs{\theta}}

\usepackage{commath}
\usepackage{amsmath,amssymb}
\usepackage{mathtools}
\usepackage{algorithm,algpseudocode}
\usepackage{hyperref}
\usepackage{color}

\usepackage{diagbox}
\usepackage{bm}
\usepackage{booktabs}       
\usepackage{multirow}
\usepackage{subfigure}
\usepackage[percent]{overpic}
\usepackage{caption}

\providecommand{\keywords}[1]
{\textbf{\text{Keywords: }} #1}
\newcommand{\bmx}{{\bm{x}}}
\newcommand{\bmm}{{\bm{m}}}
\newcommand{\bmphi}{{\bm{\varphi}}}

\newcommand{\bmza}{{\bm{z}_0}}
\newcommand{\bmzb}{{\bm{z}_1}}
\newcommand{\bmtheta}{{\bm{\theta}}}
\newcommand{\bmeta}{{\bm{\eta}}}

\newcommand{\Ucal}{{\mathcal{U}}}

\algnewcommand{\Inputs}[1]{%
  \State \textbf{Inputs:}
  \Statex \hspace*{\algorithmicindent}\parbox[t]{.8\linewidth}{\raggedright #1}
}
\algnewcommand{\Initialize}[1]{%
  \State \textbf{Initialize:}
  \Statex \hspace*{\algorithmicindent}\parbox[t]{.8\linewidth}{\raggedright #1}
}
\algnewcommand{\Outputs}[1]{%
  \State \textbf{Outputs:}
  \Statex \hspace*{\algorithmicindent}\parbox[t]{.8\linewidth}{\raggedright #1}
}

\title{PSP-GEN: Stochastic inversion of the Process-Structure-Property chain in materials design through deep, generative probabilistic modeling}

\author[a]{Yaohua Zang}
\author[a,b]{Phaedon-Stelios Koutsourelakis}
\affil[a]{Technical University of Munich, Professorship of Data-driven Materials Modeling, School of Engineering and Design, Boltzmannstr. 15, 85748 Garching, Germany}
\affil[b]{Munich Data Science Institute (MDSI - Core member), Garching, Germany}
\affil[ ]{\text{\{yaohua.zang, p.s.koutsourelakis\}@tum.de}}
\begin{document}

\maketitle

\begin{abstract}
Inverse material design is a cornerstone challenge in materials science, with significant applications across many industries. Traditional approaches that invert the structure-property (SP) linkage to identify microstructures with targeted properties often overlook the feasibility of production processes, leading to microstructures that may not be manufacturable. Achieving both desired properties and a realizable manufacturing procedure necessitates inverting the entire Process-Structure-Property (PSP) chain. However, this task is fraught with challenges, including stochasticity along the whole modeling chain, the high dimensionality of microstructures and process parameters, and the inherent ill-posedness of the inverse problem.
This paper proposes a novel framework, named PSP-GEN, for the goal-oriented material design that effectively addresses these challenges by modeling the entire PSP chain with a deep generative model. It employs two sets of continuous, microstructure- and property-aware, latent variables, the first of which provides a lower-dimensional representation that captures the stochastic aspects of microstructure generation, while the second is a direct link to processing parameters. This structured, low-dimensional embedding not only simplifies the handling of high-dimensional microstructure data but also facilitates the application of gradient-based optimization techniques.
The effectiveness and efficiency of this method are demonstrated in the inverse design of two-phase materials, where the objective is to design microstructures with target effective permeability. We compare state-of-the-art alternatives in challenging settings involving limited training data, target property regions for which no training data is available, and design tasks where the process parameters and microstructures have high-dimensional representations.
\end{abstract}

\keywords{Inverse Materials Design, PSP Chain, Deep Learning, Deep Generative Model}

\section{Introduction}
\label{sec:introduction}
Inverse material design is a critical challenge in modern materials science, offering unprecedented avenues for accelerating material discovery and optimization across diverse sectors \cite{executive_office_of_the_president_national_science_and_technology_council_materials_2011}. Its applications span crucial industries such as aerospace \cite{aage2017giga}, automotive \cite{zegard2016bridging}, pharmaceutical \cite{sanchez2017optimizing}, and renewable energy \cite{yu2013inverse}, where tailored materials with specific properties are paramount for technological advancement and innovation.
Central to the inverse design paradigm are the Process-Structure (PS) and Structure-Property (SP) linkages, which describe the causal links in material response \cite{olson_computational_1997,brough2017microstructure,yang2018deep,smith2016linking}. The PS linkage determines how processing parameters influence material microstructures, while the SP linkage relates these structures to material properties. Together, these linkages form the Process-Structure-Property (PSP) chain, providing a holistic framework for understanding and manipulating material behavior \cite{khosravani2017development,kouraytem2021modeling}.
Inverse materials design based on the PSP linkage entails identifying the processing conditions or material microstructures that yield desired properties, thereby enabling targeted material design and optimization \cite{mcdowell_integrated_2009}.
By leveraging the comprehensive understanding of the PSP linkage, inverse design methodologies empower researchers to navigate the vast design space effectively and expedite the discovery of materials with tailored properties for diverse applications \cite{bostanabad2018computational,honarmandi2022accelerated,saunders2023metal}.
However, navigating inverse materials design based on the entire PSP linkage faces many difficult challenges.
Firstly, the intricate transformation from process parameters to material microstructure constitutes a complex, stochastic, hierarchical system, characterized by high nonlinearity \cite{niezgoda2011understanding,niezgoda2013novel}. Its inherent stochasticity amplifies the complexity and challenges of inverse design. 
Secondly, the representation of several material microstructures is inherently discrete and high-dimensional \cite{damewood_representations_2023}, which complicates computational handling and renders derivative-based methods impossible due to the inability to calculate derivatives with respect to discrete-valued inputs \cite{gomez-bombarelli_automatic_2018,generale2023bayesian}.
Additionally, calculating the properties of microstructures often involves solving partial differential equations (PDEs), which are inherently time-consuming and computationally intensive \cite{smith2016linking}.
Collectively, these factors engender an inverse problem that is inherently ill-posed, stochastic, and computationally intensive, necessitating innovative methodologies and advanced computational tools to surmount these challenges effectively.

Due to these difficulties, pertinent methods frequently focus only on inverting the SP linkage (or microstructure-centered design).
Traditional methods \cite{guest2006optimizing,guest2007design,fullwood2010microstructure,andreassen2014design} for inverting the SP linkage generally include optimization algorithms, statistical modeling, and physics-based simulations.
While these methods offer more interpretability, they often suffer from slow convergence, especially in high-dimensional spaces, and struggle with the curse of dimensionality, limiting their scalability to complex material systems.
Conversely, deep-learning-based methods, particularly Deep Generative Models such as Variational Autoencoders (VAEs) \cite{gomez-bombarelli_automatic_2018,jung_microstructure_2020,kim2021exploration,kim2021exploration,xu2022harnessing,attari2023towards}, Generative Adversarial Networks (GANs) \cite{tan2020deep,lee_fast_2021} and diffusion models \cite{lyu2024microstructure}, have gained prominence for their ability to map complex and irregular microstructure space into a well-structured and continuous latent space. This latent space can then be used to establish  the requisite, differentiable  
 link to material properties which derivative-based search algorithms can explore for discovering 
 microstructures exhibiting desired properties. However, they may suffer from limitations such as data efficiency, interpretability, and generalization to unseen material systems.
Despite the advancements achieved by inverting the SP linkage, these methods only address part of the problem as they do not provide any manufacturing routes for producing these microstructures. The microstructures obtained through these methods may face the risk of being unable to find corresponding processing parameters, rendering them unfeasible to generate. Therefore, they require a secondary inversion process, such as iterative searches relying on domain knowledge.  
More crucially, these methods overlook the randomness inherent in the generation process of microstructures.

Different from the microstructure-centered design, the goal-oriented material design considers the entire PSP chain, aiming to identify optimal processing parameters that yield microstructures exhibiting desired properties \cite{tran_solving_2021}.
In goal-oriented material design, certain methodologies are categorized as microstructure-agnostic approaches, as they circumvent the intricate representation of material microstructures. Instead, they concentrate on establishing direct correlations between processing parameters and material properties, aiming to optimize the former by inverting the Process-Property (PP) linkage \cite{wei2020natural,devaraj2016low,li2016metastable}. While these techniques alleviate the computational complexities associated with irregular and high-dimensional microstructure spaces, they overlook crucial information contained in the  microstructures, essential for bridging processing parameters and properties. Consequently, they exhibit diminished performance compared to methods that explicitly incorporate microstructural information throughout the PSP linkage, especially in cases where the properties exhibit significant variability due to changes in processing parameters.
In contrast, microstructure-aware methodologies encompass not only material properties and processing conditions within the design space but also microstructural information which plays a crucial role in the solution of the optimization problem.
In \cite{molkeri2022importance}, the authors proposed a multi-fidelity method for identifying the combination of chemistry and processing parameters that  maximizes the targeted mechanical property of a model, dual-phase steel. They found that the deliberate incorporation of microstructural information into materials design  dramatically improves materials optimization. However, they did not incorporate the full microstructure
seamlessly into the design process and  instead used a feature of the microstructure, i.e. the volume fraction of the martensite phase.
Recently, the authors in \cite{generale2023bayesian,generale2024inverse} proposed a Bayesian framework leveraging a lightweight, normalizing-flow-based generative approach for the stochastic inversion of the complete PSP linkage. They employed dimension-reduced,  2-point spatial correlations as representative features of the microstructure, and a conditional, continuous normalizing flow model was employed to capture the conditional density of processing parameters given properties.

In this paper, we introduce a novel, integrated framework for goal-oriented materials design, called PSP-GEN, which effectively addresses the challenges posed by the PSP chain in its entirety. The core of PSP-GEN is a deep generative model designed to capture the full PSP linkage, enabling direct exploration of optimal solutions within the processing parameter space.
Our generative model maps PSP data to a well-structured, low-dimensional latent space, which consists of two sets of microstructure- and property-aware variables. The first part accounts for generators that are directly related to the processing variables whereas the second to the independent,  stochastic elements. Together, these representations form a comprehensive microstructural descriptor capable of reconstructing input microstructures and predicting their corresponding properties.
These continuous, latent variables not only enable the lower-dimensional embedding of the high-dimensional microstructural space and the surrogation of the expensive PS and SP links but more importantly,  the computation of derivatives which are essential in the back-propagation of information from the properties to the processing variables and especially when the latter are higher-dimensional \cite{pfeifer_process_2018}.

Hence, we can reformulate the goal-oriented design problem as a straightforward optimization problem, easily solvable with gradient-based methods. As demonstrated in our study, the proposed model significantly outperforms a recently developed state-of-the-art method in a 2D dual-phase microstructure design problem.
In summary, the main contributions of the paper are:
\bi
\item An integrated, data-driven surrogate for the whole PSP chain in contrast to multi-component strategies or methods that neglect the importance of processing in achieving not only optimal but also realizable designs. %
\item A fully probabilistic framework that accounts for the multitude of uncertainties in the PSP chain, as well as in the formulation of the inverse design problem.
\item The proposed framework simplifies the goal-oriented design problem into an optimization problem solvable with gradient-based methods even when the original microstructural representations are discrete-valued.
\ei
The remainder of the paper is structured as follows. In section \ref{sec:probdef} we provide a concrete definition of the main quantities involved, enumerate the various sources of uncertainty, and reformulate the inverse design problem in a manner that reflects the inherent stochasticity. More importantly, we discuss the computational challenges that we attempt to overcome with the proposed generative model presented in section \ref{sec:methodology}. In particular, we present the latent variables involved, and how these relate to the processing variables as well as generate both microstructures and properties. We discuss a Variational-Bayesian EM scheme \cite{beal_variational_2003} that is used for inference and learning, and more importantly, how the trained model can efficiently lead to solutions of the stochastic inverse design problem. Finally, in section \ref{sec:result}, we present comparative results on the inverse design of two-phase, heterogeneous materials where we explore the ability of our model to operate under limited data as well as to generalize to unseen property domains. We present a concluding summary as well as tangible avenues for potential extensions in section \ref{sec:conclusion}. 

\section{Inverting the  PSP chain: a formidable computational challenge }
\label{sec:probdef}
\begin{figure}[tb]
\centering
\includegraphics[width=0.75\textwidth]{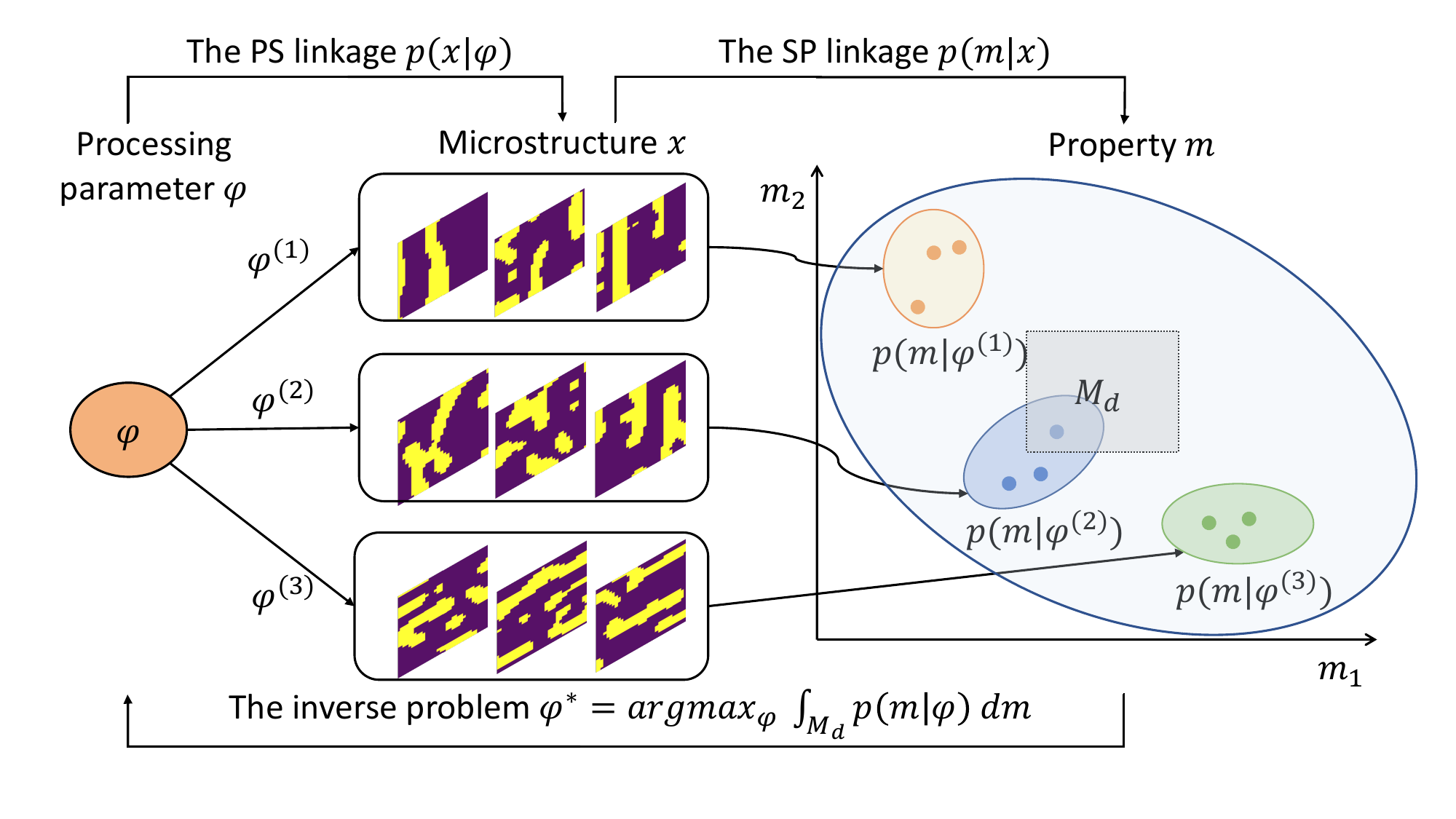}
\caption{Inverse materials design based on the whole PSP chain: the forward problem involves the PSP chain that contains two successive processes, i.e., the PS linkage $p(\bm{x}|\bm{\varphi})$ and the SP linkage $p(\bm{m}|\bm{x})$. The PS linkage is a stochastic process to generate microstructures $\bm{x}$ given processing parameters $\bm{\varphi}$. The SP linkage calculates the properties $\bm{m}$ of given microstructures $\bm{x}$, which requires the solving of complex PDEs.
The inverse design problem aims to find the optimal $\bm{\varphi}^*$ that {\em maximizes the probability that the generated microstructures possess properties falling within a target region $M_d$, i.e., $\bm{\varphi}^* = \arg\max_{\bm{\varphi}\in\mathcal{U}_{\bm{\varphi}}} \int_{M_d} p(\bm{m}|\bm{\varphi}) ~d\bm{m}$}.}
\label{fig:inv_design_problem}
\end{figure}

In this paper, we focus on the design of heterogeneous media with an emphasis on two-phase materials. Such materials are encountered in  a multitude of engineering applications such as aligned and chopped fiber composites,  dual-phase  steels,  porous membranes, particulate composites, cellular solids, colloids, microemulsions, concrete
\cite{torquato_random_2002}. In most instances, pertinent microstructures can be characterized only statistically which is essential in  determining and ultimately controlling their effectiveness e.g. mechanical, transport  properties are of great importance \cite{torquato_optimal_2010}.
In the following, we denote with:
\begin{itemize}
\item $\bm{\varphi} \in \mathcal{U}_{\bm{\varphi}}\subset \mathbb{R}^{d_{\varphi}}$ the controllable processing parameters that give rise to a family of microstructures. These serve as the optimization variables in inverse materials design.
\item $\bm{x} \in \mathcal{X}$ the microstructural representation, which for the two-phase media considered takes the form of a high-dimensional (i.e. $dim(\bx)=d_x>>1$), binary vector i.e. $\mathcal{X} \equiv \{0,1\}^{d_x}$. Multi-phase materials can similarly be represented with a discrete-valued vector. 
\item $\bs{m} \in \mathcal{M} \subset \RR^{d_{m}}$ the microstructural  properties that are targeted in the context of inverse design and  pertain  to e.g. effective permeability, elastic moduli, thermal conductivity, fracture toughness, electrical conductivity, or combinations thereof.
\end{itemize}
In Figure \ref{fig:inv_design_problem}, we elucidate the entirety of the PSP chain, depicting the sequential transformation from processing parameters $\bm{\varphi}$ to the eventual effective properties $\bm{m}$.
Most research efforts in capturing the nonlinear and multiscale processes involved in forward and backward modeling of the process-structure (PS) and structure-property (SP) linkages have disregarded uncertainties \cite{arroyave_systems_2019}. The latter is present in all steps of material analysis and design \cite{chernatynskiy_uncertainty_2013,honarmandi_uncertainty_2020}. Firstly, process variables do not uniquely determine the resulting microstructure which can exhibit aleatoric fluctuations in their features.  They rather  define a probability distribution on microstructures  \cite{popova_process-structure_2017, xu_harnessing_2022}. 
 Furthermore, the experimental data that are used to model  process-structure (most often) and  structure-property relations are contaminated by random noise and are frequently incomplete \cite{bock_review_2019}. Very often,  the models employed for the process-structure or structure-property links are  inherently stochastic and additional uncertainty exists with regard to their parameter values  or even form, especially in multiscale formulations \cite{panchal_key_2013}. Finally, the processes of  model compression and dimension reduction employed to gain computational efficiency or insight, very often lead to some  information loss  which in turn gives rise to predictive uncertainty \cite{grigo_bayesian_2019}. (Back-)propagating uncertainty through complex and potentially multiscale models poses significant computational difficulties \cite{zabaras_scalable_2008}.
Data-based surrogates offer a powerful alternative but unlike classical machine-learning applications where one operates in abundance of data. In materials problems, however, due to the cost of the experimental or computational generation procedures, pertinent tools must operate in the small-data regime, which can itself give rise to  additional (epistemic)  uncertainty. 
We note that problem formulations based on Bayesian Optimization  \cite{frazier_bayesian_2016,zhang_bayesian_2020,jung_microstructure_2020}, apart from their inherent limitation to low-dimensional design spaces, account for  uncertainty in the objective solely  due to the imprecision of the surrogate and not due to the aleatoric, stochastic variability of the underlying microstructure.

Given the aforementioned uncertainties and without loss of generality, we denote the PS and SP  links with two densities, namely $p(\bm{x}|\bm{\varphi})$ and $p(\bm{m}|\bm{x})$ respectively.
In cases where any of these links is (assumed to be) deterministic, the corresponding density can assume a degenerate, Dirac-delta form, e.g. $p(\bm{m}|\bm{x})=\delta(\bm{m}-\bm{m}(\bx))$ where  $\bm{m}(\bx)$ denotes the deterministic SP map or $p(\bm{x}|\bm{\varphi})=\delta(\bm{x}-\bm{x}(\bm{\varphi}))$ where  $\bx(\bm{\varphi})$ denotes the deterministic PS map.

The goal-oriented, inverse design seeks to find the optimal process parameters, say $\bm{\varphi}^*$, those yield microstructures exhibiting desired properties, e.g. they lie in a prescribed  target  region $M_d \subset \mathcal{M}$, as illustrated in Figure \ref{fig:inv_design_problem}.
Given the stochastic nature of the PSP chain, we reformulate the objective to identify the process parameters $\bm{\varphi}^*$ that {\em maximize the probability that the generated microstructures possess properties  within the target region $M_d$} \cite{ikebata_bayesian_2017,rixner_self-supervised_2022}, i.e.:
\begin{equation}
\label{eq:inv_problem}
\begin{array}{ll}
\bm{\varphi}^* & = \arg\max_{\bm{\varphi}\in\mathcal{U}_{\bm{\varphi}}}P\left( \bs{m} \in M_d | \bm{\varphi} \right) \\
& =
\arg\max_{\bm{\varphi}\in\mathcal{U}_{\bm{\varphi}}} \int_{M_d} p(\bm{m}|\bm{\varphi}) ~d\bm{m} \\
& = \arg\max_{\bm{\varphi}\in\mathcal{U}_{\bm{\varphi}}} \int_{M_d} \sum_{\bx \in \mathcal{X}} \underbrace{p(\bm{m}|\bm{x})}_{SP-link} ~\underbrace{ p(\bm{x}|\bm{\varphi}) }_{PS-link}~d\bm{m}. 
\end{array}
\end{equation}

Solving the optimization problem of  \refeqp{eq:inv_problem} is challenging due to the complexities of the whole PSP chain.
Firstly, the PS linkage maps an, in general, modest-dimensional space $\mathcal{U}_{\bm{\varphi}}$, into a high-dimensional, discrete space $\mathcal{X}$.
Consequently, computing the derivatives of the microstructure $\bm{x}$ with respect to the processing parameter $\bm{\varphi}$ is  impossible, rendering derivative-based methods unsuitable for solving the inverse problem directly. Additionally, the uncertainty introduced by the PS process further complicates the inversion.
Secondly, for each evaluation of  the SP link where physics-based models  (e.g.  PDEs) are used to capture the relation between microstructure and properties, solving them repeatedly can be  time-consuming, especially when the dimensionality of $\bm{x}$ is very high.
Derivatives of the properties $\bs{m}$ with respect to the discrete-valued input $\bx$ are not defined which precludes back-propagation and search in the high-dimensional microstructural space.
Moreover, multiple microstructures may share the same or similar properties, making the inverse problem highly ill-posed.
It becomes apparent therefore that,  not only computationally efficient surrogates for the aforementioned probabilistic links  are required, but more importantly, formulations that enable the computation of derivatives, an essential feature of any  optimization scheme.
This in turn necessitates continuous and preferably lower-dimensional embeddings of the discrete-valued microstructures. Classical (non)linear dimensionality reduction techniques have been commonly employed but these are not necessarily well-suited to the general problem as they are agnostic to the properties as well  as to the processing variables that we ultimately seek to optimize.

\section{Proposed Methodology}
\label{sec:methodology}
\subsection{Overview of the proposed framework}
In this paper, we propose a data-driven, deep, {\em generative} model called PSP-GEN, which captures the intricate PSP relationships and addresses the aforementioned challenges in a unified and integrated manner in contrast to the multi-component, modular strategies that have mostly been proposed thus far \cite{kalidindi_hierarchical_2015}.
It makes use of $L$ triads of data from the whole PSP chain i.e. triplets of processing parameter values $\bm{\varphi}^{(l)}$, microstructures $\bm{x}^{(l)}$ generated from those   and corresponding properties $\bm{m}^{(l)}$, where $l=1,\ldots, L$ which we denote summarily with $\mathcal{D}=\{\bm{\varphi}^{(l)}, \bm{x}^{(l)}, \bm{m}^{(l)}\}^L_{l=1}$.
We note that it is possible to have {\em unlabelled}  data i.e. process-microstructure pairs $\mathcal{D}_{PS}=\{ \bm{\varphi}^{(l)},\bx^{(l)} \}_{l=1}^{L_{PS}}$.
Such data is less expensive to obtain as one does not need to run the pertinent experiment/model for predicting properties and therefore more abundant. We discuss at the end of section \ref{sec:inference} how the proposed model can seamlessly accommodate such cases and learn from partial data as well in a semi-supervised fashion \cite{rixner_probabilistic_2021}.

To address challenges posed by microstructure space's irregular and high-dimensional nature, numerous studies have endeavored to reduce their complexity by representing images as low-dimensional objects. Various features have been proposed to aggregate local information into global descriptors, including texture statistics \cite{haralick1973textural}, volume fractions \cite{molkeri2022importance}, grain size distributions \cite{backman2006icme}, visual bag of words \cite{decost2015computer}, and auto- and cross-correlation functions \cite{fullwood2008microstructure}. An overview of these techniques is provided in \cite{larmuseau2020compact}.
Such traditional microstructural metrics may not sufficiently account for many properties of interest due to the inherent complexity of microstructures.
Deep-neural-network-based models possess greater expressivity and among these, the Variational Auto-Encoder (VAE \cite{kingma2013auto}) has  demonstrated remarkable efficacy compared with traditional microstructural metrics \cite{attari2023towards}. However, existing approaches utilizing VAEs are predominantly confined to microstructure-centered materials design and employ latent representations that are property- and process-parameter-agnostic.

\begin{figure}[tb]
\centering
\includegraphics[width=0.75\textwidth]{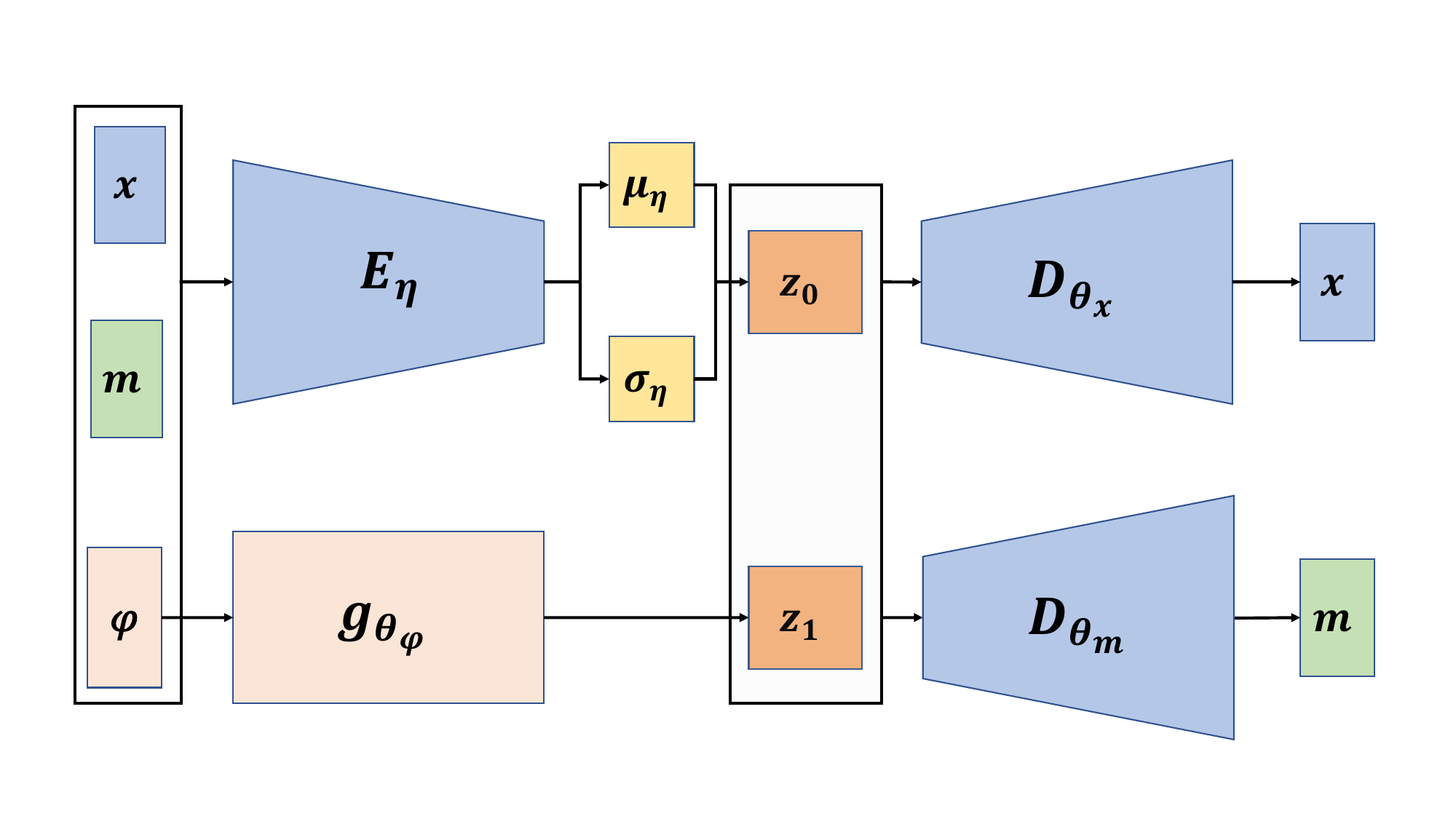}
\caption{The overview of the PSP-GEN model: the model based on the assumption $p_{\bm{\theta}}(\bm{x},\bm{m}|\bm{\varphi}) = \int p_{\bm{\theta}_\bmx}(\bm{x}|\bm{z}_0, \bm{z}_1)p_{\bm{\theta}_\bmm}(\bm{m}|\bm{z}_0, \bm{z}_1)p_{\bm{\theta}_\bmphi}(\bm{z}_1|\bm{\varphi})p(\bm{z}_0) ~d\bm{z}_1d\bm{z}_0 $ where $p(\bm{z}_0)$ is a fixed prior distribution and $p_{\bm{\theta}_\bmphi}(\bm{z}_1|\bm{\varphi})$ is a delta distribution, i.e., $\bm{z}_1$ is determined by $\bm{\varphi}$ through a neural network $g_{\bm{\theta}_\bmphi}$. Given the processing parameter $\bm{\varphi}$, the generative entails first computing $\bm{z}_1=g_{\bm{\theta}_\bmphi}(\bm{\varphi})$ and sampling  $\bm{z}_0$  from $p(\bm{z}_0)$. Subsequently  microstructures $\bx$ are generated through the decoder $\bm{D}_{\bm{\theta}_\bmx}$ which corresponds to the density $p_{\bm{\theta}_\bmx}(\bm{x}|\bm{z}_0,\bm{z}_1)$ and properties $\bs{m}$ through the decoder $\bm{D}_{\bm{\theta}_\bmm}$ which corresponds to the density $p_{\bm{\theta}_\bmm}(\bm{m}|\bm{z}_0,\bm{z}_1)$.  The Figure also depicts the encoder $\bmza=E_\bmeta(\bmphi, \bmx,\bmm)$ which corresponds to the density $q_{\bm{\eta}}(\bm{z}_0|\bmphi, \bm{x},\bm{m})$ employed during training by the proposed VB-EM scheme (section \ref{sec:inference}). 
}
\label{fig:cVAE_PSP}
\end{figure}

The proposed PSP-GEN model aims to emulate the real PSP process in generating new microstructures $\bm{x}$ and predicting their corresponding properties $\bm{m}$ given the process parameters $\bm{\varphi}$. A schematic overview of the PSP-GEN model and the inference process (section \ref{sec:inference}) is shown in Figure \ref{fig:cVAE_PSP}.
Unlike current methodologies that model the SP linkage and PS linkage separately, our model integrates the generation of microstructures and the determination in a comprehensive manner. It is based on the joint density:
\begin{equation}\label{eq:psp_model}
    p_{\bm{\theta}}(\bm{x},\bm{m}|\bm{\varphi}) =
    \int p_{\bm{\theta}_\bmx}(\bm{x}|\bm{z}_0, \bm{z}_1)p_{\bm{\theta}_\bmm}(\bm{m}|\bm{z}_0, \bm{z}_1)p_{\bm{\theta}_\bmphi}(\bm{z}_1|\bm{\varphi})p(\bm{z}_0) ~d\bm{z}_1 d\bm{z}_0 ,
\end{equation}
where $\bm{\theta}=(\bmtheta_\bmx, \bmtheta_\bmm,\bmtheta_\bmphi)$ summarily represents the model parameters.

At the core of the PSP-GEN lie the {\em continuous}, latent variables $\bm{z}=(\bm{z}_0, \bm{z}_1)$ which simultaneously provide a  lower-dimensional embedding of $\bx$ and $\bs{m}$, i.e. they are property-aware, as well as the link to the process variables $\bm{\varphi}$. In particular, knowledge of $\bz$ enables one to reconstruct both the microstructure $\bx$ through the decoding density $p_{\bm{\theta}_\bmx}(\bm{x}|\bm{z}_0, \bm{z}_1)$ and of  the properties $\bs{m}$ through $p_{\bm{\theta}_\bmm}(\bm{m}|\bm{z}_0, \bm{z}_1)$. It is this  feature that is essential to solving the inverse design problem in \refeqp{eq:inv_problem} as it will be demonstrated in the sequel. As one can deduce from \refeqp{eq:psp_model}, the  $\bz_1$ components represent the unobserved generators that are affected by the processing variables $\bm{\varphi}$ as dictated by the learnable density $p_{\bm{\theta}_\bmphi}(\bm{z}_1|\bm{\varphi})$, whereas   $\bz_0$ the remaining generators needed, which are  independent of processing. Before embarking on the presentation of the form of the densities involved (section \ref{sec:specs}) the inference of the latent variables and the learning of model parameters $\bt$ (section \ref{sec:inference}), we briefly illustrate the advantages of the PSP-GEN architecture in solving the inverse design problem i.e. \refeqp{eq:inv_problem}.

We note that the product of densities $p(\bx|\bs{\varphi})$ for the PS-link and $p(\bs{m}|\bx)$ for the SP-link define the joint $p(\bs{m},\bx|\bm{\varphi})$ which can be substituted by the learned density $p_{\bm{\theta}}(\bm{x},\bm{m}|\bm{\varphi})$ in \refeqp{eq:psp_model} of our model. In particular:

\be
\begin{array}{ll}
P\left( \bs{m} \in M_d | \bm{\varphi} \right) & =\int_{M_d} \sum_{\bx \in \mathcal{X}} p(\bm{m}|\bm{x}) ~p(\bm{x}|\bm{\varphi}) ~d\bm{m} \\
&= \int_{M_d} \sum_{\bx \in \mathcal{X}} p(\bs{m},\bx|\bm{\varphi}) ~d\bmm \\
& \approx \int_{M_d} \sum_{\bx \in \mathcal{X}}  p_{\bm{\theta}}(\bm{x},\bm{m}|\bm{\varphi})~d\bm{m} \\
& = \int_{M_d} \sum_{\bx \in \mathcal{X}} \left( \int p_{\bm{\theta}_\bmx}(\bm{x}|\bm{z}_0, \bm{z}_1)p_{\bm{\theta}_\bmm}(\bm{m}|\bm{z}_0, \bm{z}_1)p_{\bm{\theta}_\bmphi}(\bm{z}_1|\bm{\varphi})p(\bm{z}_0) ~d\bm{z}_1 d\bm{z}_0 \right)~d\bm{m} \\
& = \int_{M_d}  \left( \int p_{\bm{\theta}_\bmm}(\bm{m}|\bm{z}_0, \bm{z}_1)p_{\bm{\theta}_\bmphi}(\bm{z}_1|\bm{\varphi})p(\bm{z}_0) ~d\bm{z}_1 d\bm{z}_0 \right)~d\bm{m}
\end{array}
\label{eq:opti_redux}
\ee
At first glance, it appears that one simply exchanged the summation with respect to $\bx$ with an integration with respect to $\bz_0$ and $\bz_1$. As we demonstrate in detail in section \ref{sec:opti}, the key difference is that the densities involved are differentiable since $\bz_0$ and $\bz_1$ are continuous and one can circumvent the need to sum over the possible microstructures. We note that this does not hinge on any particular form of the associated densities but merely on the generative model structure adopted.

\subsection{Model specification}
\label{sec:specs}
In this section, we specify the densities involved in the model definition of \refeqp{eq:psp_model}, provide additional insight on  the latent variables $\bz$ as well as details on the model parameters $\bt$.
Given the role of $\bz_0 \in \RR^{d_{z_0}}$ as stochastic generators, we  employ a standard multivariate normal for the respective density,  i.e.:
\be
p(\bz_0)=\mathcal{N}(\bz_0|\bs{0,~I})
\ee
With regards to the latent variables $\bz_1 \in \RR^{d_{z_1}}$ which are assumed to be controlled by the processing variables $\bm{\varphi}$, we postulate a deterministic dependence, i.e. $\bm{z}_1=g_{\bm{\theta}_\bmphi} (\bm{\varphi})$ which means that $p_{\bm{\theta}_\bmphi}(\bm{z}_1|\bm{\varphi})$ will attain the form of a degenerate Dirac-delta density:
\begin{equation}\label{eq:latent_z1}
 p_{\bm{\theta}_\bmphi}(\bm{z}_1|\bm{\varphi})=\delta(\bm{z}_1-g_{\bm{\theta}_\bmphi} (\bm{\varphi}) ).
\end{equation}
The function $g_{\bm{\theta}_\bmphi}$ is expressed with   a neural network parameterized with   $\bm{\theta}_\bmphi$ (see  \ref{apdx:model_structure}).
The deterministic relation simplifies the aforementioned integrals as integration with respect to $\bz_1$ can be eliminated and all occurrences of $\bz_1$ can be substituted with $g_{\bm{\theta}_\bmphi} (\bm{\varphi})$. Equation \ref{eq:psp_model} will become:
\begin{equation}\label{eq:psp_model_simple}
    p_{\bm{\theta}}(\bm{x},\bm{m}|\bm{\varphi}) =
    \int p_{\bm{\theta}_\bmx}(\bm{x}|\bm{z}_0, g_{\bm{\theta}_\bmphi} (\bm{\varphi}))p_{\bm{\theta}_\bmm}(\bm{m}|\bm{z}_0, g_{\bm{\theta}_\bmphi} (\bm{\varphi})) ~p(\bm{z}_0) ~d\bm{z}_0.
\end{equation}
and \refeqp{eq:opti_redux}:
\be
P\left( \bs{m} \in M_d | \bm{\varphi}  \right) \approx  \int_{M_d}  \left( \int p_{\bm{\theta}_\bmm}(\bm{m}|\bm{z}_0, g_{\bm{\theta}_\bmphi} (\bm{\varphi})) ~p(\bm{z}_0) ~ d\bm{z}_0
\right)~d\bm{m}
\label{eq:opti_redux_simple}
\ee

Given the binary nature of the microstructural representation $\bx=\{x_j\}_{j=1}^{d_x}$, we employ the following decoding density $p_{\bm{\theta}_\bmx}(\bm{x}|\bm{z}_0,\bm{z}_1)$ \cite{tipping_probabilistic_1998,jiang2016variational}:
\begin{equation}
p_{\bmtheta_\bmx}(\bm{x}|\bm{z}_0,\bm{z}_1)=\prod_{j=1}^{d_\bmx} \sigma(\mu_{j,\bmtheta_\bmx}(\bmza, \bmzb))^{x_j}~
(1-\sigma(\mu_{j,\bmtheta_\bmx}(\bmza, \bmzb) ) )^{(1-x_j)}
\label{eq:xdecode}
\end{equation}
where $\sigma(.)$ is the sigmoid function and $\bm{\mu}_{\bmtheta_\bmx}=\{ \mu_{j,\bmtheta_\bmx}(\bmza, \bmzb) \}_{j=1}^{d_\bmx}$ are represented with  a neural network parameterized with  $\bmtheta_\bmx$ (details in  \ref{apdx:model_structure})\footnote{The model can readily be extended to multi-phase media by employing the softmax function.}

For the final decoding density for the properties $\bm{m} \in \mathcal{M} \subset \RR^{d_m}$ we employ a multivariate Gaussian of the form:
\be
p_{\bm{\theta}_\bmm}(\bm{m}|\bm{z}_0,\bz_1)=\mathcal{N}
(\bm{m}|\bm{\mu}_{\bm{\theta}_\bmm}(\bm{z}_0,\bm{z}_1), \textrm{diag}(\bm{\sigma}_{\bmtheta_\bmm}(\bmza,\bmzb)))
\label{eq:mdecode}
\end{equation}
where \(\bm{\mu}_{\bmtheta\bmm}\) and \(\bm{\sigma}_{\bmtheta\bmm}\) represent the mean and (diagonal) covariance, respectively, which  are expressed by a neural network with parameters \(\bmtheta_\bmm\) (details in  \ref{apdx:model_structure}).

\subsection{Inference and Learning }
\label{sec:inference}
We consider first the canonical scenario where triads of data $\mathcal{D}$=$\{\bm{\varphi}^{(l)}, \bm{x}^{(l)}, \bm{m}^{(l)}\}^L_{l=1}$ are available for training and discuss other cases in the sequel.
We propose learning the optimal values for the model parameters $\bt$ by maximizing the log-likelihood $\sum_{l=1}^L \log p_{\bt} (\bm{m}^{(l)}, \bm{x}^{(l)} | \bm{\varphi}^{(l)})$ which is generally analytically intractable. To this end, we advocate the use of the Variational-Bayesian Expectation-Maximization (VB-EM) scheme \cite{beal_variational_2003}, which, following \refeqp{eq:psp_model_simple} employs Jensen's inequality to construct a lower bound $\mathcal{F}$ as:
\be
\begin{array}{lll}
 & \sum_{l=1}^L \log p_{\bt}(\bm{m}^{(l)}, \bm{x}^{(l)} | \bm{\varphi}^{(l)}) 
 &  \\
 &   =    \sum_{l=1}^L   \log \int p_{\bm{\theta}_\bmx}(\bm{x}^{(l)}|\bm{z}_0^{(l)}, g_{\bm{\theta}_\bmphi} (\bm{\varphi}^{(l)}))p_{\bm{\theta}_\bmm}(\bm{m}^{(l)}|\bm{z}_0^{(l)}, g_{\bm{\theta}_\bmphi} (\bm{\varphi}^{(l)})) ~p(\bm{z}_0^{(l)}) ~d\bm{z}_0^{(l)} & \\
&   \ge \sum_{l=1}^L  \left< \log \frac{p_{\bm{\theta}_\bmx}(\bm{x}^{(l)}|\bm{z}_0^{(l)}, g_{\bm{\theta}_\bmphi} (\bm{\varphi}^{(l)})) p_{\bm{\theta}_\bmm}(\bm{m}^{(l)}|\bm{z}_0^{(l)}, g_{\bm{\theta}_\bmphi} (\bm{\varphi}^{(l)})) ~p(\bm{z}_0^{(l)}) }{q_{\bs{\eta}}(\bm{z}_0^{(l)}) }  \right>_{q_{\bs{\eta}}(\bm{z}_0^{(l)} )} &  \\
 &  = \sum_{l=1}^L \mathcal{L}^{(l)}(\bt,\bs{\eta})=\mathcal{F}(\bt,\bs{\eta}) & 
\end{array}
\label{eq:llb}
\ee
The derivation makes use of the auxiliary densities $q_{\bs{\eta}}(\bm{z}_0^{(l)} )$ parametrized by $\bs{\eta}$. It can be readily shown \cite{beal_variational_2003} that the optimal $q_{\bs{\eta}}(\bm{z}_0^{(l)} )$ coincide with the exact  posterior(s) $p_{\bt}(\bm{z}_0^{(l)}| \bm{m}^{(l)}, \bm{x}^{(l)},  \bm{\varphi}^{(l)}  )$ of the latent variables $\bm{z}_0^{(l)}$. VB-EM
alternates between maximizing $\mathcal{F}$ with respect to $\bs{\eta}$ while keeping $\bt$ fixed (VB-M step) and subsequently maximizing $\mathcal{F}$ with respect to $\bt$ while keeping $\bs{\eta}$ fixed (VB-E step). The VB-M-step seeks to find the best possible approximation to the exact posteriors within the parametrized family $q_{\bs{\eta}}$ while the VB-E-step attempts to find the optimal model parameter values given the approximate posterior.

We propose employing the following amortized form for $q_{\eta}$:
\be
q_{\eta}(\bz_0 |\bm{x},\bm{m},\bmphi)= \mathcal{N}(\bm{z}_0|\bm{\mu}_{\bm{\eta}}(\bm{x},\bm{m},\bmphi),\textrm{diag}(\bm{\sigma}_{\bm{\eta}}(\bm{x}, \bm{m}, \bm{\bmphi}))
\label{eq:z0encoder}
\ee
We note that this depends explicitly on $\bx$, $\bm{m}$, and $\bmphi$ which implies that the parameters $\bs{\eta}$ are common for all data-triads and associated latent variables $\bz_0^{(i)}$ (i.e. independent of the number of data-triads $L$).
In the aforementioned expression, $\bm{\mu}_{\bm{\eta}}$ and $\bm{\sigma}_{\bm{\eta}}$ are expressed by neural networks with parameters $\bm{\eta}$ (see   \ref{apdx:model_structure}).

Using the form of the associated densities from section \ref{sec:specs} we can simplify the expressions for the lower-bounds $\mathcal{L}^{(l)}(\bt,\bs{\eta})$ in \refeqp{eq:llb}. In particular, if we drop the data index $(l)$ which appears as a superscript in the expressions, we obtain:
\be
\begin{array}{ll}
\mathcal{L}(\bt,\bs{\eta}) & =
\left< \log p_{\bm{\theta}_\bmx}(\bm{x}|\bm{z}_0, g_{\bm{\theta}_\bmphi} (\bm{\varphi})) \right>_{q_{\bs{\eta}}(\bm{z}_0 ) }
+ \left< \log p_{\bm{\theta}_\bmm}(\bm{m}|\bm{z}_0, g_{\bm{\theta}_\bmphi} (\bm{\varphi})) \right>_{q_{\bs{\eta}}(\bm{z}_0 )} \\
 & + \left< \log \frac{ p(\bm{z}_0) }{q_{\bs{\eta}}(\bm{z}_0)}  \right>_{q_{\bs{\eta}}(\bm{z}_0 )} \\
& = \mathcal{L}_{REC,\bm{x}}(\bt,\bs{\eta})+\mathcal{L}_{REC,\bm{m}}(\bt,\bs{\eta})+\mathcal{L}_{KL}(\bs{\eta}).
\end{array}
\label{eq:elboi}
\ee
The first term attempts to minimize the reconstruction error of the microstructural  decoder, i.e. from \refeqp{eq:xdecode} and for $\bz_1=g_{\bm{\theta}_\bmphi} (\bm{\varphi})$:
\be
\mathcal{L}_{REC,\bm{x}}(\bt,\bs{\eta})= \sum_{j=1}^{d_\bmx}~ x_j \left< \log  \sigma(\mu_{j,\bmtheta_\bmx}(\bmza, \bmzb)) \right>_{q_{\bs{\eta}}}+(1-x_j) \left<
\log(1-\sigma(\mu_{j,\bmtheta_\bmx}(\bmza, \bmzb) ) \right>_{q_{\bs{\eta}} }.
\ee
The second term pertains to the  reconstruction error of  the property decoder, i.e from \refeqp{eq:mdecode} and for $\bz_1=g_{\bm{\theta}_\bmphi} (\bm{\varphi})$:
\be
\mathcal{L}_{REC,\bm{m}}(\bt,\bs{\eta})= -\frac{1}{2} \left< \sum_{j=1}^{d_\bmm} \log \sigma_{j,\bmtheta_\bmm}(\bmza,\bmzb) \right>_{q_{\bs{\eta}} }-\frac{1}{2} \left< \|\bmm-\bm{\mu}_{\bmtheta_\bmm}(\bmza,\bmzb)\|^2_{\bm{\sigma}_{\bmtheta_\bmm}(\bmza,\bmzb)} \right>_{q_{\bs{\eta}} }.
\ee
Finally, the third term acts as a regularizer by minimizing the KL-divergence between the approximate posterior $q_{\bs{\eta}}(\bz_0)$ and the prior $p(\bz_0)$. It can be simplified to:
\be
\mathcal{L}_{KL}(\bs{\eta})=
-\frac{1}{2}\left( \|\bm{\mu}_{\bm{\eta}}( \bm{x},\bm{m}, \bmphi)\|^2_2 +\|  \bm{\sigma}_{\bm{\eta}}( \bm{x}, \bm{m}, \bmphi) \|^2_2- \sum_{j=1}^{d_{\bm{z}_0}} \log \sigma_{j,\bm{\eta}}(\bm{x},\bm{m}, \bmphi)\right).
\ee
To compute the expectations with respect to $q_{\bs{\eta}}$ that appear in $\mathcal{L}$ (as well as in its gradient), we make use of the reparameterization trick \cite{kingma2013auto} by drawing $\bz_0$-samples according to (see \refeqp{eq:z0encoder}):
\be
\bz_0=\bm{\mu}_{\bm{\eta}}(\bm{x},\bm{m}, \bmphi)+\bm{\sigma}_{\bm{\eta}}(\bm{x}, \bm{m}, \bmphi) \odot \bs{\epsilon}, \qquad \bs{\epsilon}\sim \mathcal{N}(\bs{0,I}).
\label{eq:reparam_trick}
\ee
where $\odot$ denotes the Hadamard product and $\bm{\epsilon}$ is a random variable which does not depend on $\bm{\eta}$.

The gradient computation of the ELBO $\mathcal{L}$ concerning the model parameters $(\bmtheta,\bmeta)$ is carried out using automatic differentiation in PyTorch \cite{paszke2017automatic}. A schematic illustration is provided in Figure \ref{fig:cVAE_PSP}
 and a pseudocode in Algorithm \ref{alg:foward}.
For the stochastic gradient ascent needed for the maximization of $\mathcal{F}$ with respect to $\bt$ and $\bs{\eta}$, we employed the   ADAM scheme, the hyperparameters of which are reported in section \ref{sec:result}.
\begin{algorithm}[!t]
\caption{The training of the PSP-GEN}\label{alg:foward}
\begin{algorithmic}
\Inputs{$\mathcal{D} = \{\bmphi^{(l)}, \bmx^{(l)}, \bmm^{(l)}\}^{L}_{l=1}$, The dimensions $d_\bmza$ and $d_\bmzb$, The parameter $\beta$}
\Initialize{Model parameters $\bmtheta=(\bmtheta_\bmx, \bmtheta_\bmm, \bmtheta_\bmphi)$ and $\bmeta$, Learning rate $lr$.}
\While{Converged or maximum iterations reached}
\State {Obtain $\bmzb^{(l)}=g_{\bmtheta_\bmphi}(\bmphi^{(l)})$ and
$$
\bmza^{(l)}=\bm{\mu}_{\bm{\eta}}(\bmphi^{(l)}, \bmx^{(l)}, \bmm^{(l)})+\bm{\sigma}_{\bm{\eta}}(\bmphi^{(l)}, \bmx^{(l)}, \bmm^{(l)}) \odot \bs{\epsilon}, \qquad \bs{\epsilon}\sim \mathcal{N}(\bs{0,I}).
$$}
\State {Calculate the lower-bounds $\mathcal{L}^{(l)}(\bmtheta,\bmeta)$ using Equation \refeqp{eq:elboi}
and the ELBO $\mathcal{F}(\bmtheta,\bmeta)$.}
\State \textbf{VB-M step:} {Update parameter $\bmeta$ with SGD while keeping $\bmtheta$ fixed:
$$\bmeta \leftarrow \bmeta + lr \odot \nabla_{\bmeta} \mathcal{F}(\bmtheta,\bmeta).$$}
\State \textbf{VB-E step:} {Update parameter $\bmtheta$ with SGD while keeping $\bmeta$ fixed:
$$\bmtheta \leftarrow \bmtheta + lr \odot \nabla_{\bmtheta} \mathcal{F}(\bmtheta,\bmeta).$$ }
\If{Validation error does not decrease within 20 epochs}
\State {Update the learning rate to $lr \leftarrow lr/3$.}
\EndIf
\EndWhile
\end{algorithmic}
\end{algorithm}

\noindent \textbf{Remarks:}\\
\bi
\item to obtain disentangled latent representations for $\bm{z}_0$ and $\bm{z}_1$, one has to balance the trade-offs between the constituent  ELBO terms \cite{dai2019diagnosing}.
A state-of-the-art solution is the so-called $\beta$-VAE trick \cite{higgins2017beta} which places an appropriate weight $\beta$ on the KL-divergence term, resulting in the following learning objective (compare with \refeqp{eq:elboi}):
\be
\begin{array}{ll}
\mathcal{L}(\bt,\bs{\eta})
& = \mathcal{L}_{REC,\bm{x}}(\bt,\bs{\eta})+\mathcal{L}_{REC,\bm{m}}(\bt,\bs{\eta})+\beta~\mathcal{L}_{KL}(\bs{\eta}).
\end{array}
\label{eq:elboibeta}
\ee
The effect and calibration of the parameter $\beta$ are discussed in  \ref{apdx:beta}.

\item We briefly discuss how the proposed model can also take advantage of {\em unlabelled}, inexpensive data consisting of PS pairs, i.e.  $\mathcal{D}_{PS}=\{ \bm{\varphi}^{(l)},\bx^{(l)} \}_{l=1}^{L_{PS}}$. We note that from \refeqp{eq:psp_model_simple} one can marginalize the properties $\bs{m}$ in order to obtain the model density:
 \be
 p_{\bm{\theta}}(\bm{x}|\bm{\varphi}) =
    \int p_{\bm{\theta}_\bmx}(\bm{x}|\bm{z}_0, g_{\bm{\theta}_\bmphi} (\bm{\varphi})) ~p(\bm{z}_0) ~d\bm{z}_0.
\label{eq:psp_model_simple_partial}
\end{equation}
This can serve as the likelihood of the unlabelled data $\mathcal{D}_{PS}$ above which in turn can be added to the (log)likelihood of the full, labeled data in \refeqp{eq:llb}. The unlabelled log-likelihood can also be bounded from below by introducing an auxiliary density, say $q_{\bs{\zeta}}(\bz_0)$. which attempts to approximate the posterior $p_{\bt}(\bm{z}_0^{(l)}|  \bm{x}^{(l)},  \bm{\varphi}^{(l)})$ for each unlabelled data pair $l=1,\ldots,L_{PS}$ in $\mathcal{D}_{PS}$. The resulting sum over the data-pairs of the log-likelihood lower-bounds  will be a function of $\bt, \bs{\eta}$ and $\bs{\zeta}$ and can be maximized using the aforementioned VB-EM scheme. Although this avenue is not explored in the numerical experiments of section \ref{sec:result}, we note its significance in enabling semi-supervised training and  reducing  the dependence on the
expensive, labeled data \cite{kingma_semisupervised_2014}.
\ei

\subsection{Solving the inverse design problem}
\label{sec:opti}
In this section, we discuss  how the aforementioned generative model can be used to efficiently obtain solutions to the inverse design problem defined in
\refeqp{eq:inv_problem}. To this end and  recalling \refeqp{eq:opti_redux_simple}  for the sought probability  $P\left( \bs{m} \in M_d | \bm{\varphi} \right)$, we derive expressions for its gradient which are now possible due to the differentiability of $p_{\bm{\theta}_\bmm}(\bm{m}|\bm{z}_0, \bz_1)$ (see \refeqp{eq:mdecode}) and the fact that $\bz_0,\bz_1$ are continuous. We note that gradients are essential in order to obtain solutions efficiently and necessary when the number of process variables $\bm{\varphi}$ (i.e. the optimization variables) is high. In particular:
\be
\begin{array}{ll}
\frac{\pa P\left( \bs{m} \in M_d | \bm{\varphi} \right)  }{\pa  \bm{\varphi} } = \int_{M_d}   \int \frac{ \pa \log p_{\bm{\theta}_\bmm}  }{\pa \bz_1} \frac{\pa g_{\bm{\theta}_\bmphi} }{\pa \bm{\varphi}}
 ~p_{\bm{\theta}_\bmm}(\bm{m}|\bm{z}_0, g_{\bm{\theta}_\bmphi} (\bm{\varphi}) )p(\bm{z}_0) ~ d\bm{z}_0 ~d\bm{m}
\end{array}
\ee
The derivatives in the expression can be computed point-wise using back-propagation and the integrals involved  can be readily approximated by Monte Carlo using $M$ samples $\{ \bs{m}^{(i)}, \bz_0^{(i)} \}_{i=1}^M$ drawn from $p_{\bm{\theta}_\bmm}(\bm{m}|\bm{z}_0, g_{\bm{\theta}_\bmphi} (\bm{\varphi}) )p(\bm{z}_0)$ as:
\be
\frac{\pa P\left( \bs{m} \in M_d | \bm{\varphi} \right)  }{\pa  \bm{\varphi} } \approx \left( \frac{1}{M}
\sum_{i=1}^M  \left. 1_{M_d}(\bs{m}) \frac{ \pa \log  p_{\bm{\theta}_\bmm}(\bm{m}^{(i)}|\bm{z}_0^{(i)}, \bz_1 ) }{\pa \bz_1}  \right|_{\bz_1=g_{\bm{\theta}_\bmphi} (\bm{\varphi})} \right) \frac{\pa g_{\bm{\theta}_\bmphi} (\bm{\varphi}) }{\pa \bm{\varphi}}
\ee
where $1_{M_d}(\bs{m})$ is the indicator function of the target domain.
Alternatively, one can get rid of the indicator function by sampling $\bm{m}^{(i)}$  from  the uniform density in $M_d$ instead (while $\bz_0^{(i)}$ are sampled from $p(\bz_0)$ as before),  in which case the Monte Carlo estimator of the gradient would be:
\be
\begin{array}{ll}
\frac{\pa P\left( \bs{m} \in M_d | \bm{\varphi} \right)  }{\pa  \bm{\varphi} }
 & \approx \left( \frac{|M_d|}{M}
\sum_{i=1}^M  \left. p_{\bm{\theta}_\bmm}(\bm{m}^{(i)}|\bm{z}_0, g_{\bm{\theta}_\bmphi} (\bm{\varphi}) )  \frac{ \pa \log   p_{\bm{\theta}_\bmm}(\bm{m}^{(i)}|\bm{z}_0^{(i)}, \bz_1 ) }{\pa \bz_1}  \right|_{\bz_1=g_{\bm{\theta}_\bmphi} (\bm{\varphi})} \right) \frac{\pa g_{\bm{\theta}_\bmphi} (\bm{\varphi}) }{\pa \bm{\varphi}} \\
& = \left( \frac{|M_d|}{M}
\sum_{i=1}^M  \left. \frac{ \pa    p_{\bm{\theta}_\bmm}(\bm{m}^{(i)}|\bm{z}_0^{(i)}, \bz_1 ) }{\pa \bz_1}  \right|_{\bz_1=g_{\bm{\theta}_\bmphi} (\bm{\varphi})} \right) \frac{\pa g_{\bm{\theta}_\bmphi} (\bm{\varphi}) }{\pa \bm{\varphi}}
\end{array}\label{eq:obj_mc}
\ee
where $|M_d|$ is the hyper-volume of the target property domain.
The aforementioned estimator can then be used in conjunction with Stochastic Gradient Ascent to maximize the  probability in the objective and to identify the optimal processing parameter values $\bm{\varphi}^*$.

\noindent \textbf{Remarks:}
\bi
\item While only point estimates of the processing variables are computed, it is also possible to obtain fully Bayesian inference results of the whole posterior $p(\bm{\varphi}|\bm{m} \in M_d)$ as is done in \cite{generale2023bayesian}. The latter can provide insight into the sensitivity of the objective to variations in $\bm{\varphi}$ that reflect all the uncertainties in the PSP  model.
Given a prior $p(\bm{\varphi})$ (e.g. a uniform on $U_{\bm{\varphi}}$) and  \refeqp{eq:opti_redux_simple}, we obtain the {\em joint} posterior by a simple application of Bayes's rule:
\be
p(\bm{\varphi}, \bz_0|\bm{m} \in M_d) \propto p_{\bm{\theta}_\bmm}(\bm{m} \in M_d|\bm{z}_0, g_{\bm{\theta}_\bmphi} (\bm{\varphi})) ~p(\bm{z}_0) ~p(\bmphi)
\ee
Exact as e.g. HMC \cite{betancourt2017conceptual} or approximate as e.g. SVI \cite{hoffman2013stochastic}  inference tools can be readily employed due to the fact that all variables involved are continuous and derivatives of all the log-densities  appearing in the expression above with respect to $\bz_0,\bmphi$ can be inexpensively computed.

\item Other property-driven objectives can also be employed in order to identify the optimal processing variables $\bmphi$. In \cite{rixner_self-supervised_2022} some of those are discussed, such as the (expected) deviation of the actual properties $\bm$ from some target $\bm{m}_{target}$ i.e. $\mathbb{E}\left[ \|\bm{m} - \bm{m}_{target} \|_2^2~| \bmphi \right] $ or  the KL-divergence $KL\left( p_{target}(\bm{m})~ || ~p(\bm{m} | \bmphi) \right)$  between a desired/target property density, e.g. $p_{target}(\bm{m})$ and the actual density $p(\bm{m} | \bmphi)$. While these cases are not discussed, it can be readily shown that they can be easily accommodated within the PSP-GEN model due to its aforementioned features, i.e. continuity of the latent space and differentiability of the probabilistic decoders.
\ei

\section{Results and Discussion}
\label{sec:result}
This section contains the results of several numerical experiments performed in the context of inverse design based on the whole PSP chain. We first discuss the PS and SP links employed to generate the data used for training and validation in the ensuing examples (section \ref{sec:data_sim}). The goals of the illustrations contained are:
\bi
\item to comparatively assess the performance of the  method herein in relation to the recently proposed approach for inverting  the entire PSP chain by \cite{generale2023bayesian,generale2024inverse}.  In brief, the latter utilizes 2-point  correlation functions as sufficient microstructural features which are subsequently subjected to a  Principal Component Analysis (PCA) to reduce their dimension. 
Sparse variational multi-output Gaussian processes (SV-MOGP) \cite{schulz2018tutorial} are employed to surrogate the forward PS and SP linkages and a continuous normalizing flow (CNF) is utilized to approximate the posterior distribution of $\bm{\varphi}$ (see first remark at the end of section \ref{sec:opti}). To facilitate comparisons, we only report point estimates of the optimal processing parameter $\bm{\varphi}^*$. Since a publicly available code repository was not provided in \cite{generale2023bayesian}, the numerical results reported are based on our own implementation of this method as described in the respective paper (and with the same parameter values reported therein) which also involved the implementation of the SV-MOGP model  using the package \textit{GPyTorch} \cite{gardner2018gpytorch}.
\item to provide insight on the role of the latent variables $\bz=(\bz_0,\bz_1)$ employed by our model and their ability to provide lower-dimensional, microstructure- and property-aware embeddings (section \ref{sec:case_1}).

\item to assess the performance for various target property domains (Section \ref{sec:case_1}) and especially those located far from the training data (section \ref{sec:case_2}). We note that the latter case is most relevant in realistic design scenarios where one should suggest new configurations and not the ones already known. Simultaneously, this poses the most challenging scenario for data-driven models as they are asked to generate accurate, {\em extrapolative} predictions.

\item to assess the performance under various amounts of training data. Given the expense of experimentally or computationally simulating the PS and SP linkages in realistic settings, it is essential that data-driven methods can also operate in the  {\em small-data} regime as well as provide accurate estimates of the associated epistemic uncertainty (section \ref{sec:case_3}).

\item to assess the performance with respect to the dimension of $\bx$ i.e. the microstructural representation  (section \ref{sec:case_4}).
\item to assess the performance under various dimensions of the process variable vector i.e. of the  design space. In particular, we consider two cases where $dim(\bmphi)=2$ and $dim(\bmphi)=10$ (section \ref{sec:case_5}).
 \ei
 
With regards to the comparison method, in all experiments, we select the number of principal components (PCs) to explain $90\%$ of the variance in the training dataset. For the proposed method, we set the dimension $d_\bmzb$ of the latent space $\bmzb$ to be consistent with the number of PCs to ensure a fair comparison.
However, it is worth noting that the proposed model can still achieve good performance with a much smaller dimension, as demonstrated in Section \ref{sec:case_1}.
For the latent space $\bmza$, a higher dimension $d_\bmza$ results in better recovery of the microstructure $\bmx$ but increases the cost of training the forward model. Therefore, we set the dimension of the latent space $d_\bmza$ to approximately $5\%$ of $d_\bmx$ to balance accuracy and training effort.
\subsection{Experimental setup}
\label{sec:dataset}
\subsubsection{The forward simulation}
\label{sec:data_sim}
We focus on designing two-dimensional, dual-phase  material microstructures where one phase corresponds to voids (infinite permeability) and the other to an impervious solid (zero permeability). The properties of interest relate to the effective permeability of the microstructure. 
The particular choice is made because it corresponds to an infinite contrast ratio in the properties of the phases. As a result, the effective property (permeability) would depend on higher-order statistics of the microstructure and in particular the presence of percolating paths of the void phase \cite{torquato_random_2002}. 

For the PS process, we generate microstructures under various process parameters using a Gaussian Random Field (GRF) and a phase-field simulation approach.
Specifically, we generate images with size $36\times 36$ using the GRF with a spectral density function (SDF) defined as $S(\bm{\varphi})$, where $\bm{\varphi}$ denotes the processing parameters. 
Subsequently, we convert the pixel values to  binary by thresholding the values. The threshold selected determines the volume fraction of the phases and in our study, it was selected such that it yields a volume fraction of $2/3$ for the void phase. We use these images as the initial condition in a phase-field model based on the Cahn-Hilliard equation \cite{cahn1965phase} which generates the  final microstructures $\bm{x}$.
The Cahn-Hilliard equation, commonly used to model phase separation in binary mixtures, is defined as:
\begin{equation}\label{eq:CH_eq}
\frac{\partial c}{\partial t}=D\nabla^{2}\left(c^{3}-c-\gamma \nabla^{2}c\right),\quad t\in [0, T]
\end{equation}
where $c$ represents the concentration with $c=\pm 1$ indicating domains, $D$ denotes the diffusion coefficient, and $\sqrt{\gamma}$ determines the length of the transition regions between the domains. In our study, we set the parameters to the following values: $D=10$, $\gamma=2$, and $T=5s$.

For the SP process, i.e.  to calculate the effective permeability of microstructures, we employ the  Stokes flow  equations with periodic boundary conditions which were solved with the finite element method (FEM) \cite{vianna2020computing}. 
We compute the effective permeability in the horizontal and vertical directions i.e. $dim(\bm{m})=2$.
In order to automatically ensure the positivity of the effective properties, we operate on their logarithms, i.e. $\bm{m}=(m_1,m_2)$ represent the logarithms of the effective permeability in the horizontal and vertical direction respectively. 

\subsubsection{Data generation and model setups}
We first defined a SDF $S(\bmphi)$ depending on two parameters, i.e. $dim(\bmphi)=2$ and generated pertinent data for the numerical experiments in sections \ref{sec:case_1}, \ref{sec:case_2} and \ref{sec:case_3}\footnote{We also considered a SDF where  $dim(\bmphi)=10$ which was used for the experiments in \ref{sec:case_5} which is discussed therein.}. In particular, we employed the following SDF:
\begin{equation}
 S(\bmphi) = \varphi_1 e^{-\varphi_2 k_x^2} + (1-\varphi_1)  e^{-\varphi_2 * k_y^2},\quad \bmphi\in \mathcal{U}_\varphi=[1,9]\times[0.2,2]
\end{equation}
where $\bmphi=(\varphi_1, \varphi_2)$ denotes the processing parameters, and $k_x, k_y$ represent shifted Fourier coordinates which were taken to be $k_x, k_y = \{0,\cdots, d_{\bmx}-1\} - \frac{d_{\bmx}}{2}.$ 

We constructed three datasets, namely:
\begin{itemize}
\item \textbf{Training dataset}: We sample uniformly in $\mathcal{U}_\varphi$  $2,500$ different values of the processing parameters $\bmphi$. 
We generate $4$ microstructure samples $\bmx$ for each processing parameter value and compute their corresponding properties $\bmm$. Consequently, the training dataset comprises $N_{train}=10,000$ triads of $(\bmphi, \bmx, \bmm)$.
\item \textbf{Testing dataset:}
We sample uniformly $500$ values of the process parameters  in  $\mathcal{U}_\varphi$. We generated $4$ microstructures $\bmx$ for each process parameter value and computed their associated properties $\bmm$. Thus, the testing dataset consists of $N_{test}=2000$ triads of $(\bmphi, \bmx, \bmm)$. In Figure \ref{fig:case1_2dz} one can see the effective permeability values attained by the microstructures in this dataset.
\item \textbf{Reference dataset:}
We also generate a very large, reference dataset which we used to obtain, whenever possible,  reference solutions to the optimization problems considered as well as to assess the objective values for the optimal $\bmphi^*$ identified by the proposed and comparison method. To generate this dataset, we considered a $20\times 20$ uniform grid in $\mathcal{U}_\varphi$ for the  process parameters $\bmphi$. We then generated $500$ microstructures $\bmx$ for each process parameter value and computed their properties $\bmm$. Therefore, the  reference dataset consists of  $N_{ref}=200,000$ triads of $(\bmphi, \bmx, \bmm)$.  
\end{itemize}

For training purposes, we divided the training dataset into two parts, using $90\%$ for training and the remaining $10\%$ for validation. We set the batch size to $50$ and employ the Adam optimizer with an initial learning rate of $lr = 10^{-3}$ to update the network parameters. Additionally, we adopt a learning rate decay technique, reducing the learning rate to one-third if the validation error does not improve within $20$ epochs \cite{goyal2017accurate}. We train both models for $500$ epochs to ensure convergence.
For solving the inverse problem, we employ a total of $M = 50,000$ samples $\{\bz_0^{(i)}, \bmm^{(i)}\}$ in the  Monte Carlo estimate of the gradient of the objective  
as in \refeqp{eq:obj_mc}. We then solve the maximization problem \refeqp{eq:inv_problem} using  Stochastic Gradient Descent (SGD)  with a batch size of $100$. We apply a learning rate decay technique to reduce the learning rate to one-fourth if the loss does not decrease within $2$ epochs, with the initial learning rate set to $lr = 0.2$. The total number of epochs is set to $20$ to ensure convergence.

\subsection{Case 1: Intra-Region design}
\label{sec:case_1}
\begin{figure}[tb] \centering
\begin{overpic}[width=0.22\textwidth]{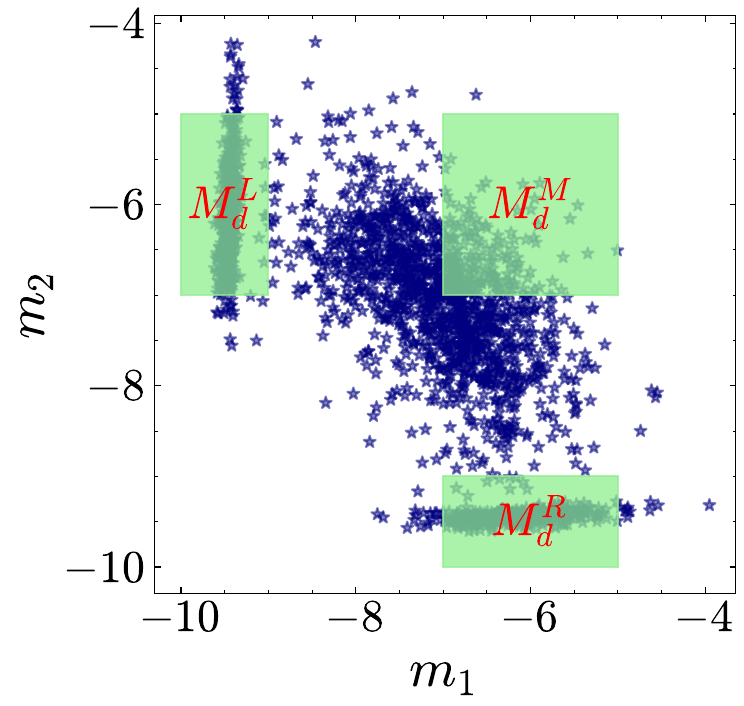}
\put(0,90){a}
\end{overpic}
\hspace{-0.05cm}
\begin{overpic}[width=0.245\textwidth]{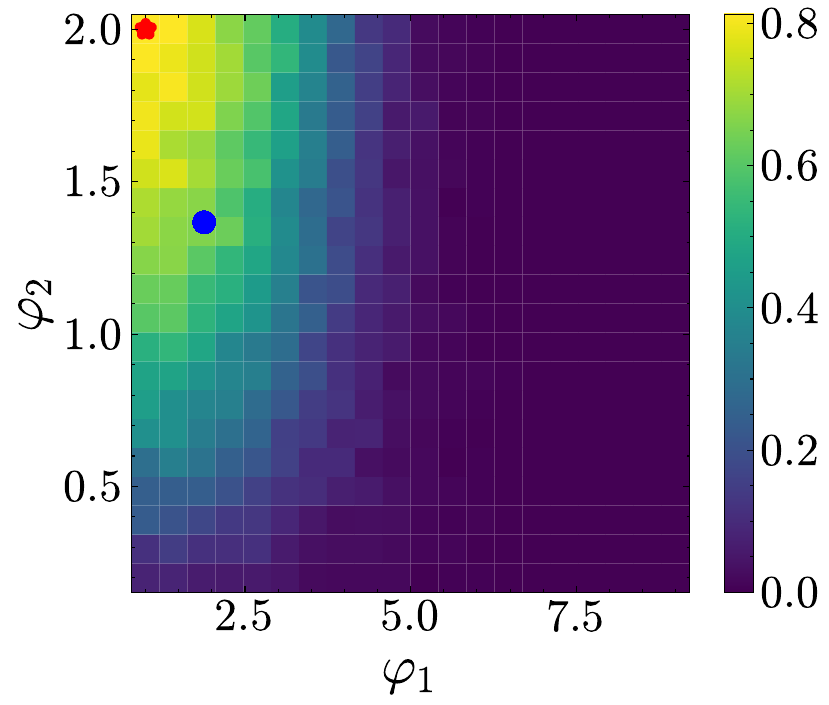}
\put(0,80){b}
\end{overpic}
\hspace{-0.05cm}
\begin{overpic}[width=0.245\textwidth]{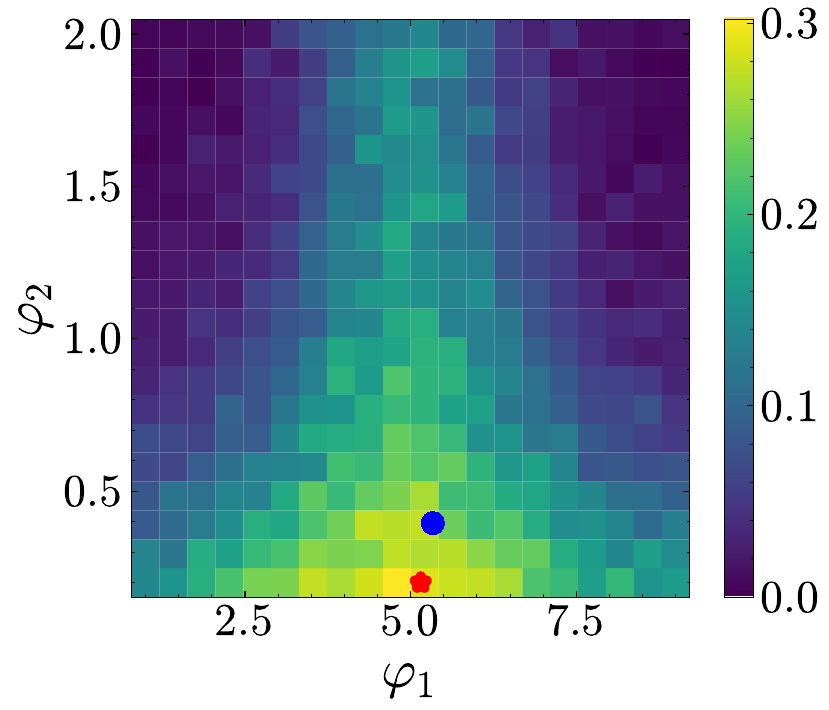}
\put(0,80){c}
\end{overpic}
\hspace{-0.05cm}
\begin{overpic}[width=0.245\textwidth]{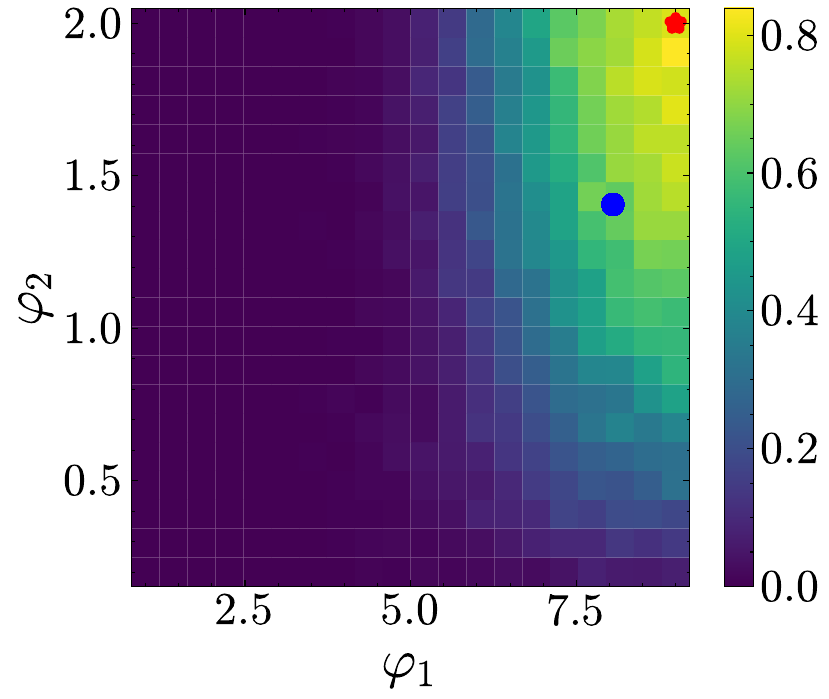}
\put(0,80){d}
\end{overpic}
\caption{The inverse design results in Case 1: (a) The regions in green indicate the three target regions $M^L_d$, $M^M_d$, and $M^R_d$, respectively. The dots represent the distribution of properties in the testing dataset. (b), (c), (d): From left to right, these figures correspond to the results for target regions $M^L_d$, $M^M_d$, and $M^R_d$, respectively. Each pixel in the images represents a processing parameter, with the pixel value indicating the percentage of microstructures generated by this processing parameter that have the desired properties (the larger the value, the closer it is to the optimal solution). The blue dot represents $\bmphi_{comp}$ obtained by the competitive method, and the red dot represents $\bmphi_{prop}$ obtained by the PSP-GEN.}
\label{fig:case1_inverse}
\end{figure}

In this section, we aim to solve the inverse materials design problem \refeqp{eq:inv_problem} for three different target regions, i.e.:
\bi
\item $M^{L}_d = [-10,-9]\times[-7,-5]$, 
\item $M^{M}_d = [-7,-5]\times[-7,-5]$, and 
\item $M^{R}_d = [-7,-5]\times[-10,-9]$
\ei
as depicted in Figure \ref{fig:case1_inverse}a.
The target regions $M^{L}_d$ and $M^{R}_d$ correspond to  low effective  permeability in the horizontal and vertical directions, respectively.
In contrast, the target region $M^{M}_d$ pertains to  high effective permeability in both the horizontal and the vertical directions.
\begin{table}
\centering
\caption{The optimal process parameters obtained by the competitive method ($\bmphi_{comp}$) and the proposed PSP-GEN ($\bmphi_{prop}$) for inverse design tasks with different target region $M_d$ in Case 1. The $\bmphi_{ref}$ denotes the reference ground truth of the processing parameter and $p_{M_d}:=p(\bmm\in M_d|\bmphi)$ represents the probability that the result properties corresponding to $\bmphi$ are located in the target region $M_d$.}
\begin{tabular}{cccc} \bottomrule
$\bmphi^*$/$p_{M_d}$ & $M^L_d$ &$ M^M_d$ &	$M^R_d$ \\ \bottomrule
$\bmphi_{comp}$/$p_{M_d}$& (1.883, 1.367)/$65.8\%$& (5.335, 0.395)/$27.4\%$& (8.052, 1.406)/$63.4\%$\\
$\bmphi_{prop}$/$p_{M_d}$& (1.009, 2.000)/$81.2\%$& (5.152, 0.200)/$28.6\%$&	(8.999, 2.000)/$79.2\%$	\\ 
$\bmphi_{ref}$/$p_{M_d}$& (1.000, 2.000)/$81.2\%$&	(4.789, 0.200)/$30.2\%$&	(9.000, 1.905)/$83.8\%$	\\ \toprule
\end{tabular}
\label{tab:case_1}
\end{table}

We denote with $\bmphi_{prop}$ the optimal process parameter values identified by the proposed PSP-GEN model (i.e. those that maximize the probability that the resulting properties would lie in each of the aforementioned target regions) and with $\bmphi_{comp}$  those found  by the competitive method. We report their values  for the aforementioned  three different target property domains in Table \ref{tab:case_1}.

In order to assess the performance of each method, 
we estimated using the reference dataset the probability that the microstructures corresponding to each $\bmphi$-grid-point considered to have properties in the target region $M_d$, denoted as $p_{M_d}:=p(\bmm\in M_d|\bmphi)$. The results for all $\bmphi$ values are depicted in Figures \ref{fig:case1_inverse}b-d where the values of $\bmphi_{prop}$ and $\bmphi_{comp}$ are indicated with a red and a blue dot respectively.
The $\bmphi$ with the highest probability is also reported in Table \ref{tab:case_1} as $\bmphi_{ref}$.
The $p_{M_d}$ values corresponding to $\bmphi_{prop}$, $\bmphi_{comp}$, and $\bmphi_{ref}$ for different target regions are recorded in Table \ref{tab:case_1}. As shown, the proposed method achieved higher $p_{M_d}$ values than the competitive method in all cases.

\begin{figure}[!t] 
\centering
\hspace{-0.05cm}
\begin{overpic}[width=0.675\textwidth]{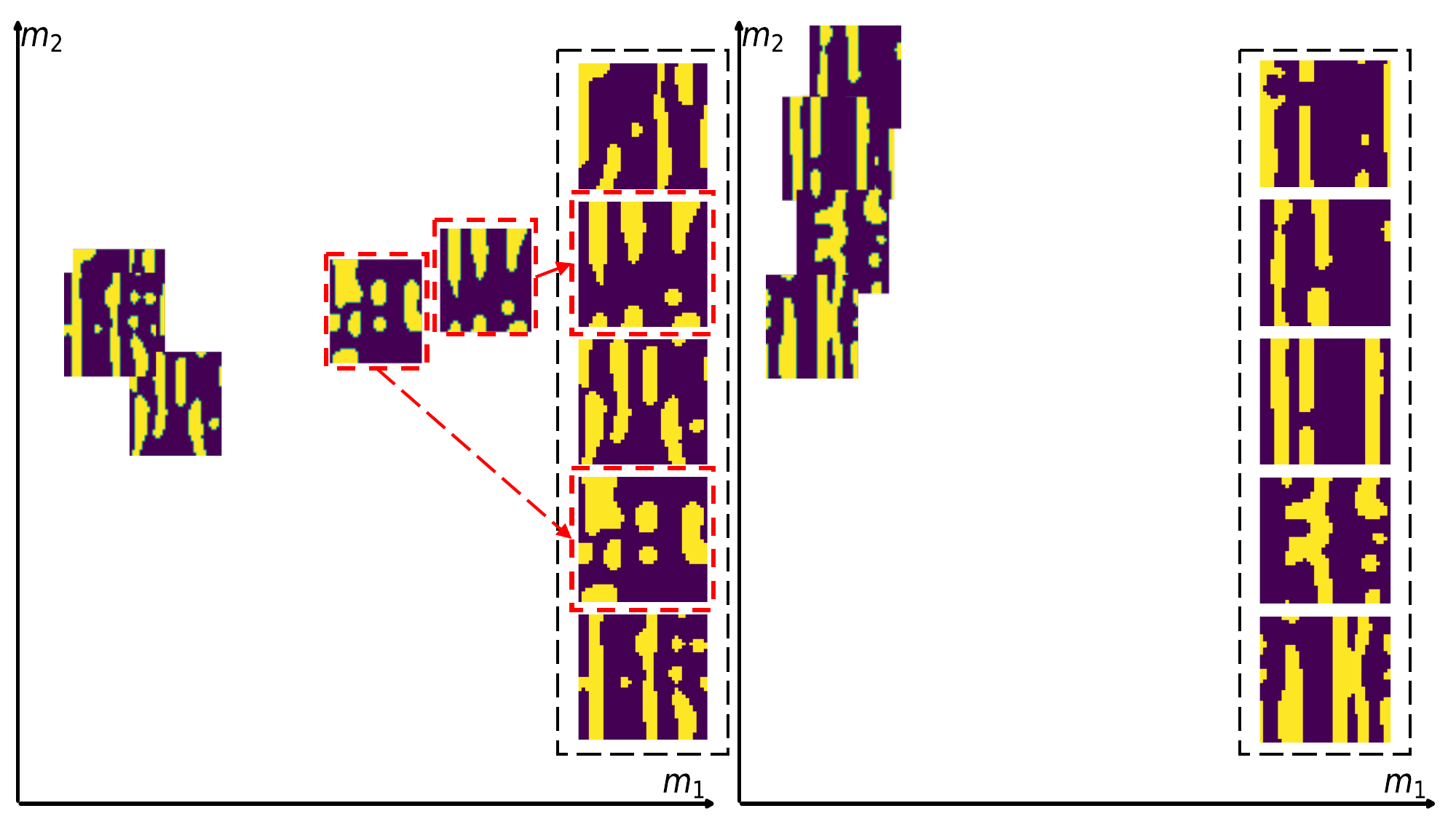}
\put(-2.5, 53){a}
\end{overpic}
\hspace{-0.05cm}
\begin{overpic}[width=0.675\textwidth]{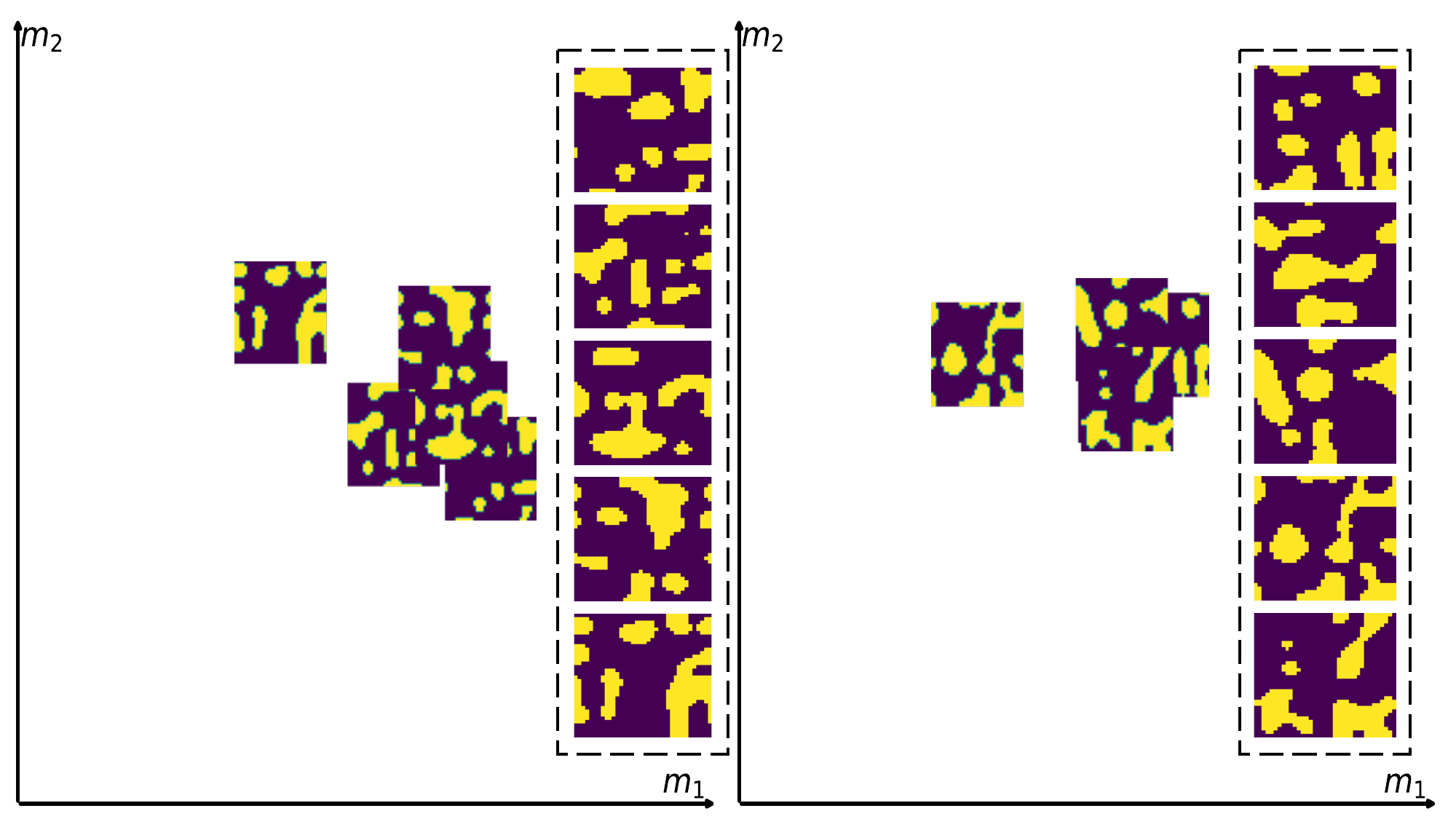}
\put(-2.5,53){b}
\end{overpic}
\hspace{-0.05cm}
\begin{overpic}[width=0.675\textwidth]{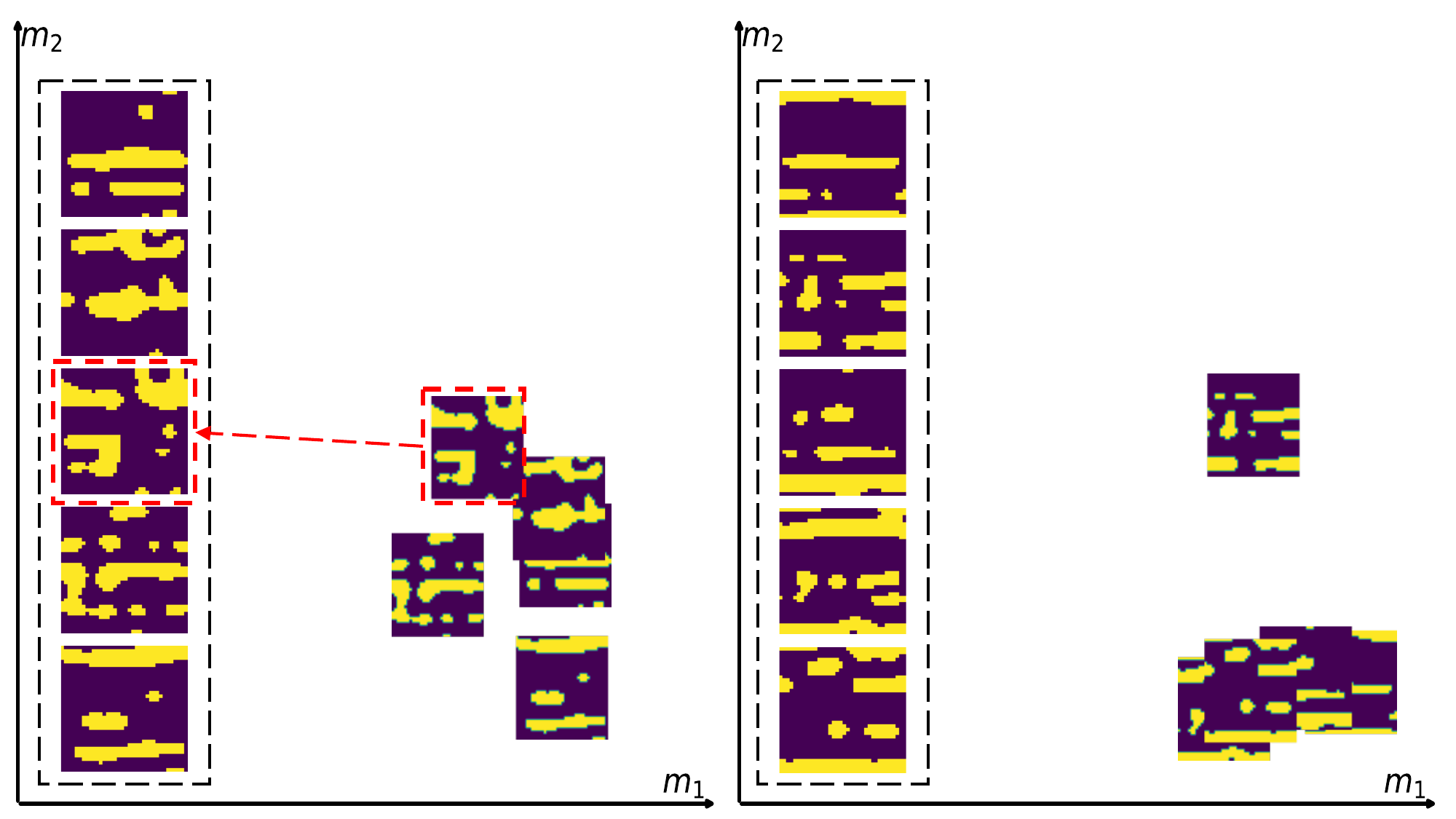}
\put(-2.5,53){c}
\end{overpic}
\caption{The microstructures generated with optimal processing parameters obtained by different methods for Case 1. (a) microstructures generated with $\bmphi_{comp}$ (left) and $\bmphi_{prop}$ (right) for target region $M^L_d$. (b) microstructures generated with $\bmphi_{comp}$ (left) and $\bmphi_{prop}$ (right) for target region $M^M_d$. (c) microstructures generated with $\bmphi_{comp}$ (left) and $\bmphi_{prop}$ (right) for target region $M^R_d$. The solid phase is indicated in yellow.}
\label{fig:case1_check_x}
\end{figure}

In addition, for each of the  three target property regions, we use the identified $\bmphi_{comp}$ and $\bmphi_{prop}$ to generate 5 microstructures, which are displayed in Figures \ref{fig:case1_check_x}a, \ref{fig:case1_check_x}b, and \ref{fig:case1_check_x}c. In each figure, the left side represents the microstructures generated by $\bmphi_{comp}$, and the right side represents the microstructures generated by $\bmphi_{prop}$.
For target region $M^L_d$, we aim to generate microstructures with low permeability in the horizontal direction, i.e. with the void phase (denoted with black) should form as little as possible connected paths in the horizontal direction. Figure \ref{fig:case1_check_x}a shows that the microstructures generated using $\bmphi_{prop}$ exhibit this characteristic, i.e. the solid phase (denoted in yellow) blocks as much as possible flow in the horizontal direction. In contrast, we observe  microstructures generated by $\bmphi_{comp}$ and indicated by the red dashed boxes that  do not meet this criterion.
Similarly, for target region $M^R_d$, we aim at microstructures where the void phase (indicated in black) forms as little as possible connected paths in the vertical direction. 
However, in Figure \ref{fig:case1_check_x}c, we notice that the microstructures within the red dashed boxes which were generated by $\bmphi_{comp}$ do not exhibit this feature, whereas the microstructures obtained using $\bmphi_{prop}$ do.

Based on the  results above, it is evident that the solution obtained by the PSP-GEN closely aligns with the ground truth computed from the reference dataset for all three regions, indicating its superior performance as compared to the competitor.
The relative inferiority  of the latter may be attributed to two main reasons. Firstly, as previously discussed, in terms of their permeability, the two phases exhibit an infinite contrast ratio, rendering the (dimension-reduced) 2-point spatial correlations inadequate to  fully capture the percolation paths that determine the effective permeabilities of the medium.  By leveraging the representation capabilities of neural network architectures and the latent variables advocated, the PSP-GEN can capture more accurately the salient microstructural features. Secondly, the innovative structure of the PSP-GEN model, which partitions the latent space into two parts, results in a more precise approximation of the PSP linkage. 

\begin{figure}[!t] \centering
\begin{overpic}[width=1.0\textwidth]{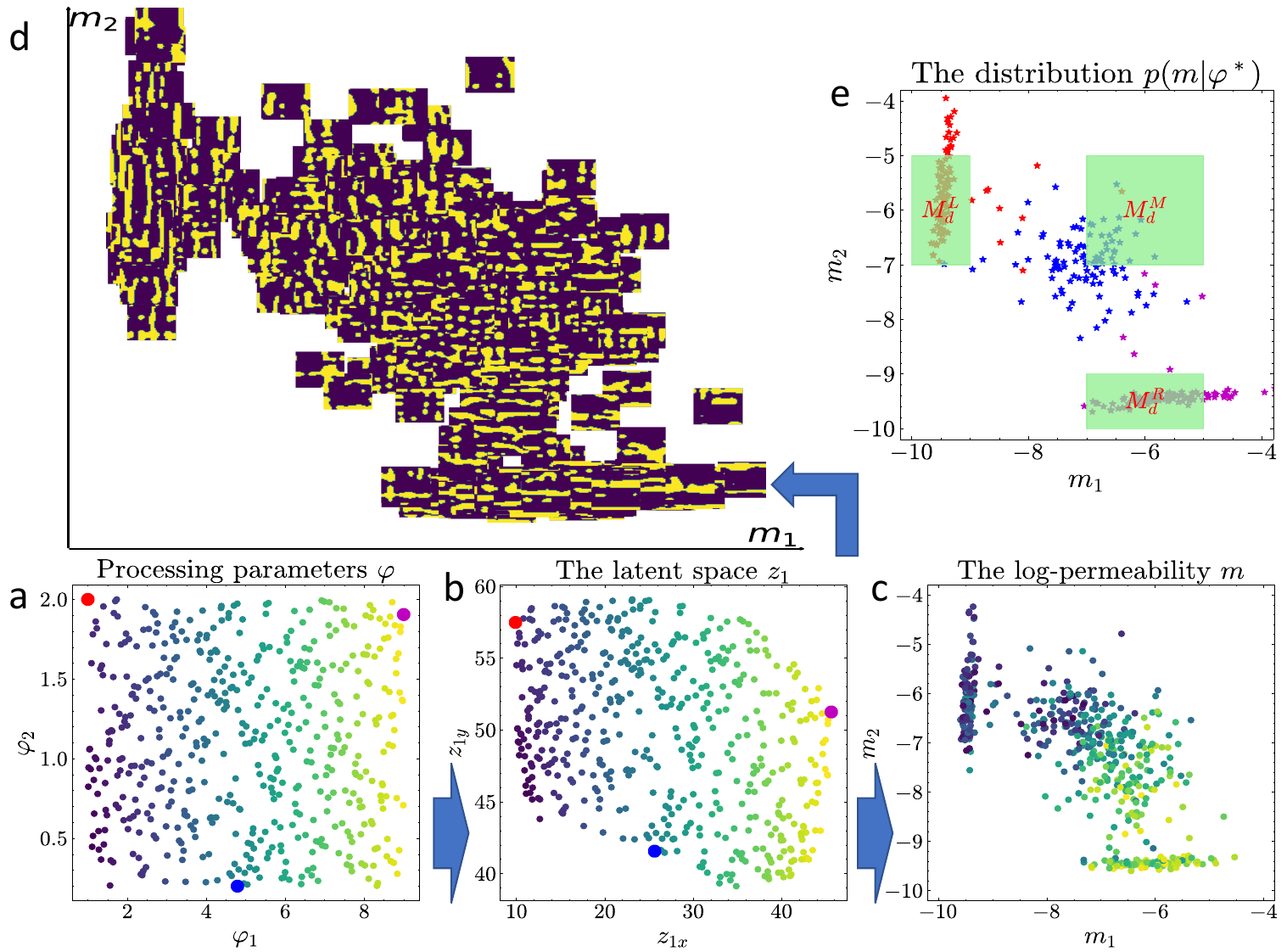}
\end{overpic}
\caption{The visualization of the forward process of the PSP-GEN model in Case 1. (a) The processing parameters $\bmphi$ in the testing dataset, with points displayed in different color shades based on their corresponding properties. The dots in red, blue, and purple represent the optimal processing parameters obtained by the PSP-GEN model for target regions $M^L_d$, $M^M_d$, and $M^R_d$, respectively. (b) The latent variables $\bmzb$ corresponding to the processing parameters $\bmphi$. The dots in red, blue, and purple represent the latent variables $\bmzb$ corresponding to the optimal processing parameters for target regions $M^L_d$, $M^M_d$, and $M^R_d$, respectively. (c) The properties $\bmm$ corresponding to the processing parameters $\bmphi$ in the testing dataset. (d) The distribution of microstructures $\bmx$, with locations determined by the corresponding properties. (e) The distribution of $p(\bmm|\bmphi^*)$, where $\bmphi^*$ represents the obtained solution for the target region $M_d\in\{M^L_d, M^M_d, M^R_d\}$. The dots in red, blue, and purple represent properties corresponding to optimal processing parameters for target regions $M^L_d$, $M^M_d$, and $M^R_d$, respectively.}
\label{fig:case1_2dz}
\end{figure}

For the same task, we also consider solving the inverse design problem using the proposed method with the dimension of the latent space $\bmzb$ set to $2$ instead of $21$ which is the number of PCs used in the competitive method, while keeping all other model parts unchanged. 
After training the model, we evaluated $\bmzb =g_{\bmtheta_\bmphi}(\bmphi)$ for each processing parameter value $\bmphi$ in the testing dataset and visualized it in Figure \ref{fig:case1_2dz}b.
In Figure \ref{fig:case1_2dz}a and \ref{fig:case1_2dz}c, we also plotted the distribution of processing parameters and the corresponding properties, respectively. Points in these Figures are displayed in different color shades based on their corresponding properties.
The distribution of microstructures $\bmx$ was visualized in Figure \ref{fig:case1_2dz}d, with the location determined by the corresponding property.
In Figures \ref{fig:case1_2dz}a-d, one readily  observes the connection between the processing parameter $\bmphi$ and the latent variable $\bmzb$, and how they influence the distribution of microstructures $\bmx$ and properties $\bmm$. This connection is vital for identifying the optimal processing parameter that leads to the desired properties.
For the target regions $M^L_d$, $M^M_d$, and $M^R_d$,  the optimal processing parameters $\bmphi_{prop}$ obtained by PSP-GEN are $(1., 2.)$, $(4.579,0.2)$, and $(8.999, 1.999)$, respectively. 
We plotted these parameters in Figure \ref{fig:case1_2dz}a with red, blue, and purple colors, respectively. The corresponding deterministic latent variables $\bmzb$ are plotted in \ref{fig:case1_2dz}b using the same colors.
To verify whether these obtained processing parameters lead to properties within the desired regions, we generated $100$ microstructures for each obtained processing parameter and calculated the corresponding properties. We visualized the distribution of $p(\bmm|\bmphi^*)$ in Figure\ref{fig:case1_2dz}e, where $\bmphi^*$ represents the obtained solution for the target region $M_d\in\{M^L_d, M^M_d, M^R_d\}$.
As shown, the distribution of properties corresponding to the identified solution in each case aligns with our expectations, falling within the target region to the greatest extent.
Compared to the reference values $\bmphi_{ref}$ in Table \ref{tab:case_1}, the PSP-GEN model still obtains good approximations despite the low dimension $d_\bmzb$ of   latent space which  further demonstrates the ability of the PSP-GEN model to extract useful, property-predicting features.

\begin{figure}[tb] \centering
\begin{overpic}[width=0.225\textwidth]{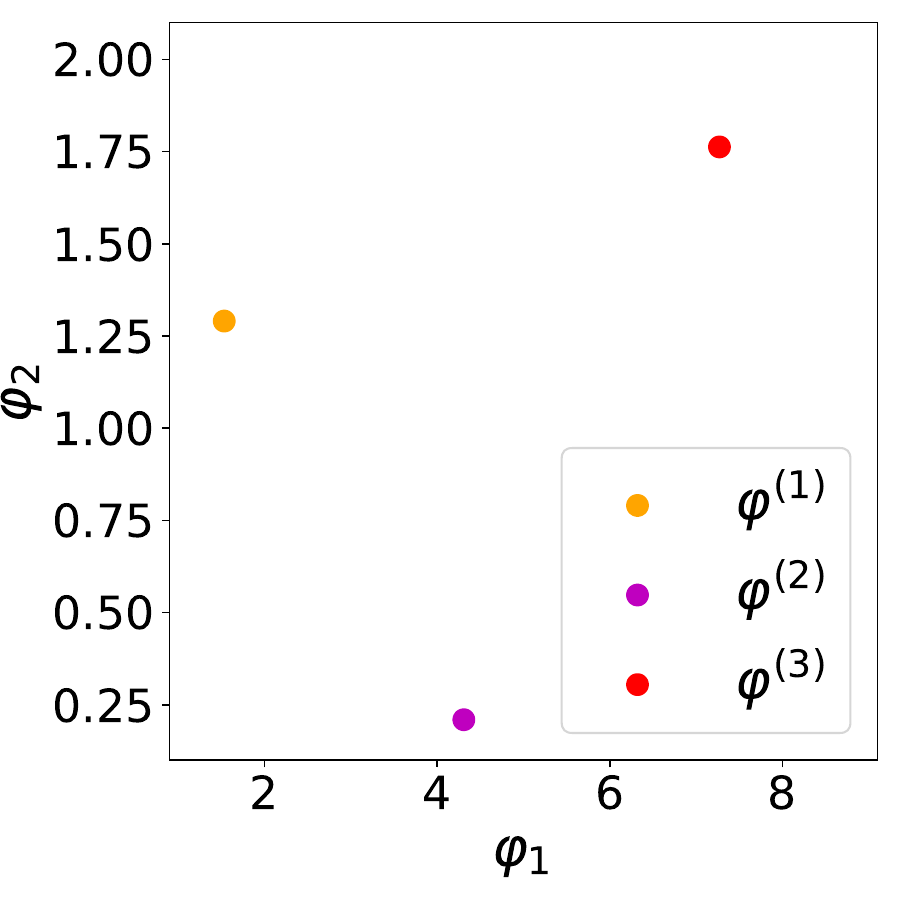}
\put(-5,95){a}
\end{overpic}
\hspace{-0.05cm}
\begin{overpic}[width=0.225\textwidth]{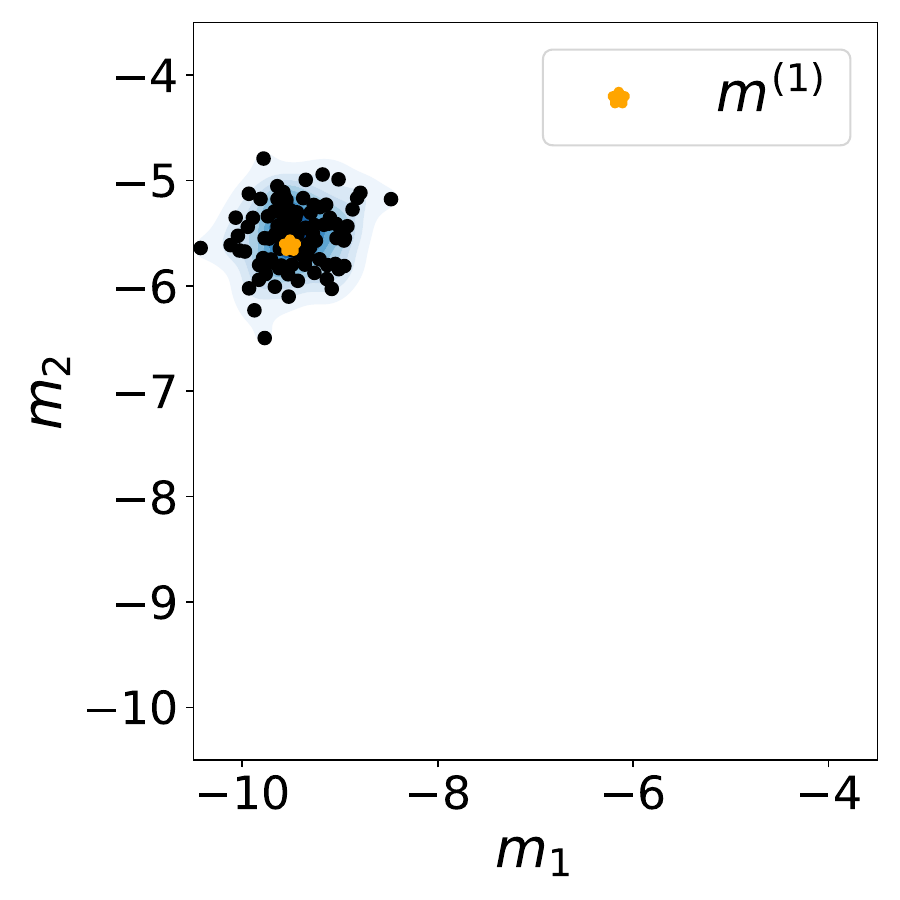}
\put(0,95){b}
\end{overpic}
\hspace{-0.05cm}
\begin{overpic}[width=0.225\textwidth]{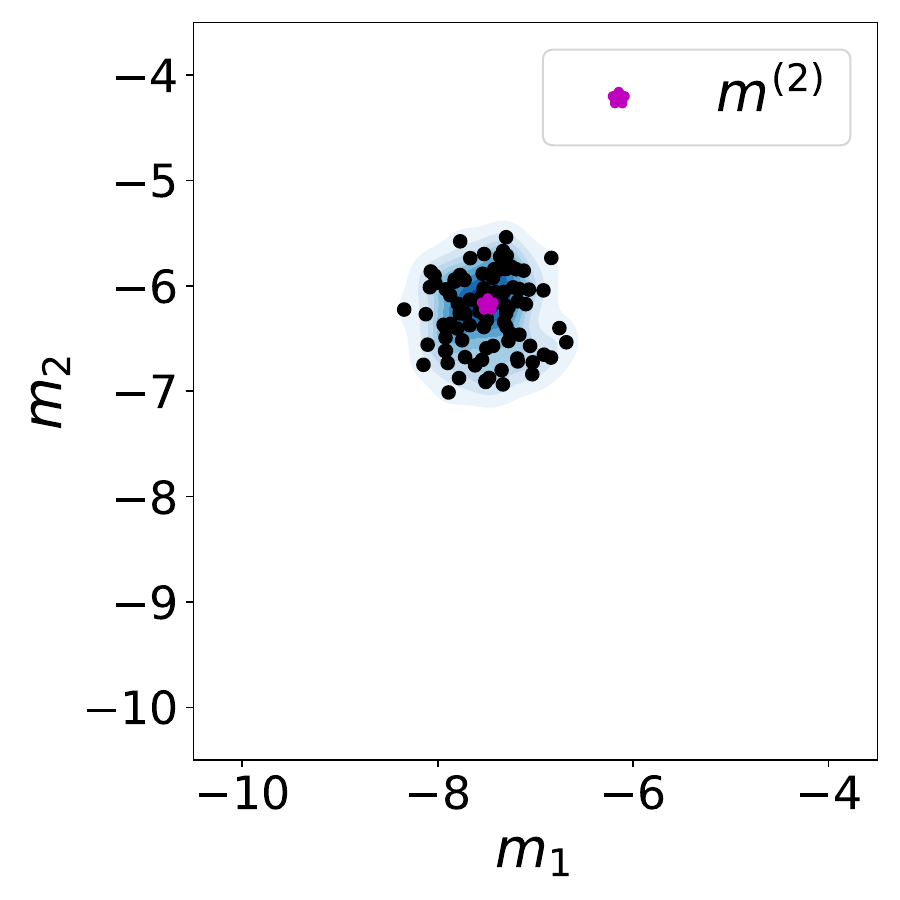}
\put(0,95){c}
\end{overpic}
\hspace{-0.05cm}
\begin{overpic}[width=0.225\textwidth]{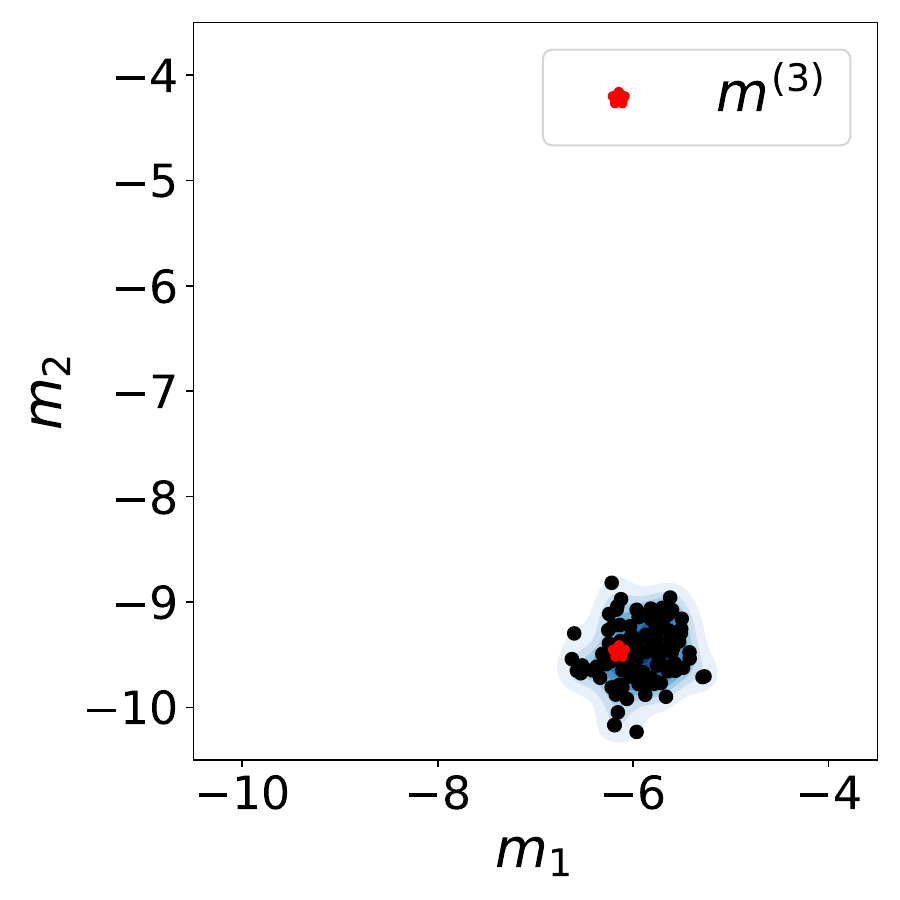}
\put(0,95){d}
\end{overpic}
\caption{The verification of the role of the latent variable $\bmza$ in the PSP-GEN model in Case 1. (a) Three processing parameters, $\bmphi^{(1)}$, $\bmphi^{(2)}$, and $\bmphi^{(3)}$, were randomly chosen within $\Ucal_{\bmphi}$. (b) The dots labeled $\bmm^{(1)}$ represent properties corresponding to $\bmx^{(1)}$, generated using the processing parameter $\bmphi^{(1)}$.
(c) The dots labeled $\bmm^{(2)}$ represent properties corresponding to $\bmx^{(2)}$, generated using the processing parameter $\bmphi^{(2)}$.
(d) The dots labeled $\bmm^{(3)}$ represent properties corresponding to $\bmx^{(3)}$, generated using the processing parameter $\bmphi^{(3)}$.
For each $i = 1, 2, 3$, the black dots represent 100 properties sampled according to the distribution $\bmm \sim p(\bmm|\bmza, g_{\bmtheta_\bmphi}(\bmphi^{(i)}))$. Here, $\bmza$ is sampled from the posterior distribution $p(\bmza|\bmx^{(i)}, \bmphi^{(i)}) \propto p(\bmx^{(i)}|\bmza, g_{\bmtheta_\bmphi}(\bmphi^{(i)})) p(\bmza)$.}
\label{fig:case1_2dz_z0}
\end{figure}

Finally, we also explored the role of $\bmza$ in the proposed framework. We first randomly sampled three $\bmphi^{(1)}, \bmphi^{(2)}, \bmphi^{(3)}$ from the space $\Ucal_{\bmphi}$, as shown in Figure \ref{fig:case1_2dz_z0}a. For each of these processing parameters, we use it to generate one microstructure and calculate the corresponding property. The properties $\bmm^{(1)}, \bmm^{(2)}, \bmm^{(3)}$ corresponding to $\bmphi^{(1)}, \bmphi^{(2)}, \bmphi^{(3)}$ are shown in Figures \ref{fig:case1_2dz_z0}b, \ref{fig:case1_2dz_z0}c, and \ref{fig:case1_2dz_z0}d, respectively.
Then, we drew $100$ samples of $\bmza$ from the posterior distribution $p(\bmza|\bmx^{(i)},\bmphi^{(i)}) \propto p(\bmx^{(i)}|\bmza,g_{\bmtheta_\bmphi}(\bmphi^{(i)})) p(\bmza)$, for each $i=1,2,3$. For each $\bmza$ sample, we drew a sample of $\bmm$ from $p(\bmm|\bmza, g_{\bmtheta_\bmphi}(\bmphi^{(i)}))$, where $i=1,2,3$. The distribution of these $\bmm$ samples is represented by black dots. For cases $i=1,2$ and $3$, they are shown in Figures\ref{fig:case1_2dz_z0}b, \ref{fig:case1_2dz_z0}c and \ref{fig:case1_2dz_z0}d. Clearly, for each $\bmm^{(i)}$, it is surrounded by the $\bmm$ samples drawn from $p(\bmm|\bmza, g_{\bmtheta_\bmphi}(\bmphi^{(i)}))$, $i=1,2,3$.

\subsection{Case 2: Cross-Region generalization}
\label{sec:case_2}
\begin{figure}[tb] \centering
\begin{overpic}[width=0.23\textwidth]{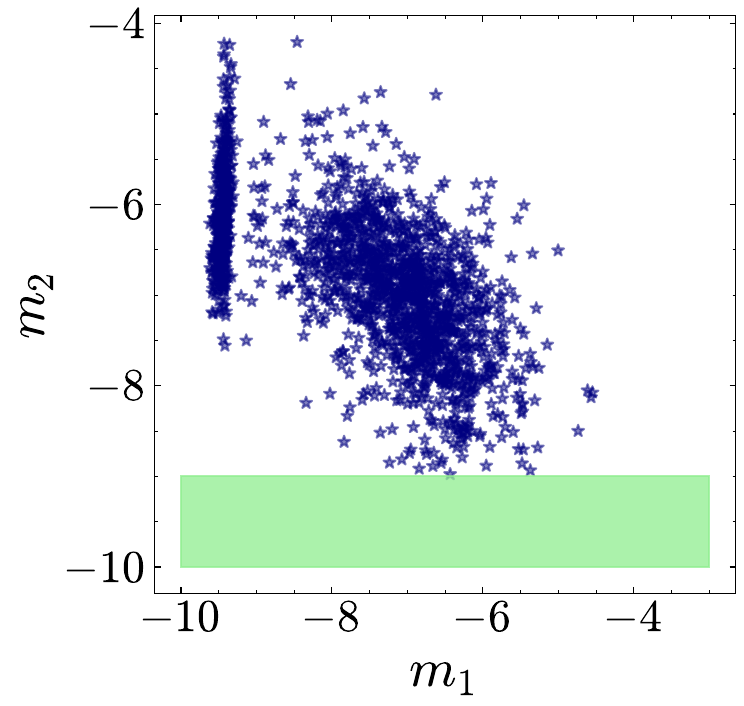}
\put(0,90){a}
\end{overpic}
\hspace{-0.05cm}
\begin{overpic}[width=0.255\textwidth]{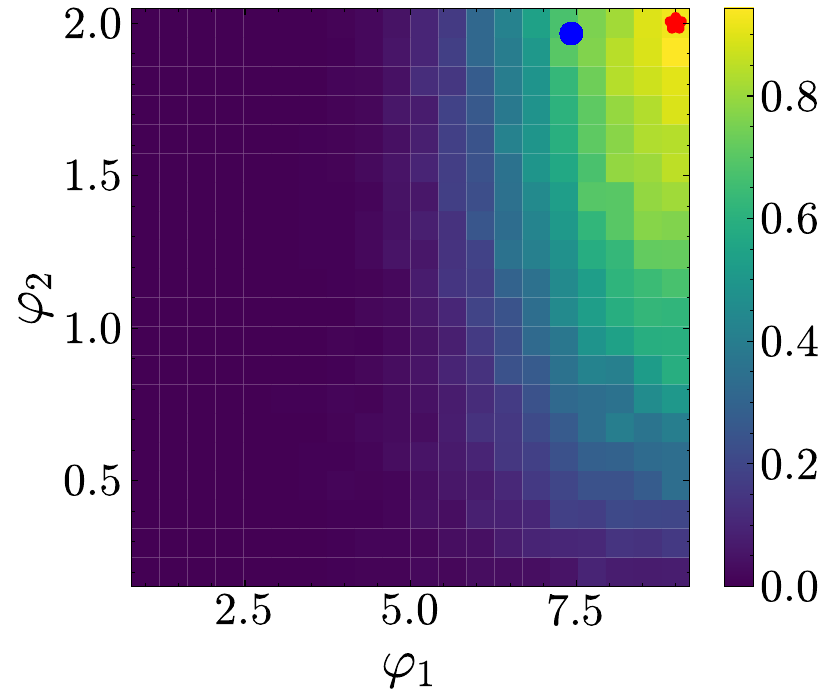}
\put(0,80){b}
\end{overpic}
\hspace{-0.05cm}
\begin{overpic}[width=0.23\textwidth]{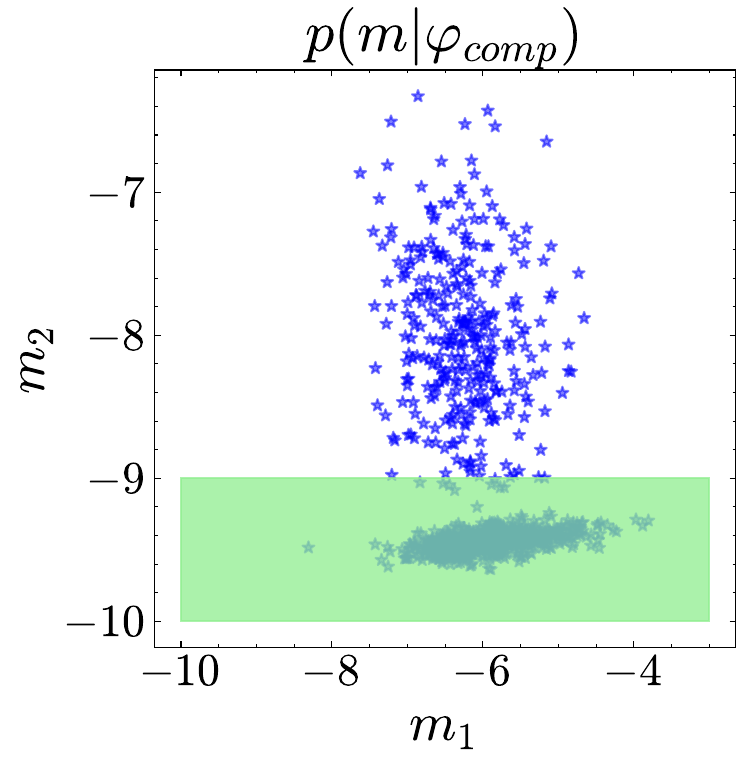}
\put(0,90){c}
\end{overpic}
\hspace{-0.05cm}
\begin{overpic}[width=0.23\textwidth]{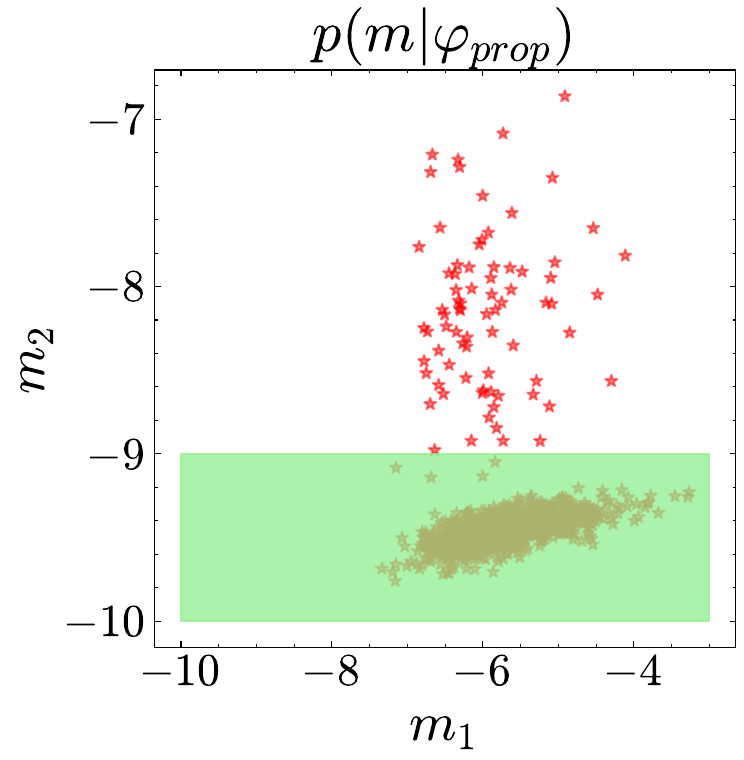}
\put(0,90){d}
\end{overpic}
\caption{The inverse design results in Case 2: (a) The green region indicates the target region $M_d$, and the dots represent the distribution of properties that lie outside of $M_d$ in the testing dataset. (b) Each pixel in this image represents a processing parameter, and the value of the pixel indicates the percentage of microstructures generated by this processing parameter that have properties within $M_d$. The blue and red dots represent $\bmphi_{comp}$ obtained by the competitive method and $\bmphi_{prop}$ obtained by the proposed PSP-GEN, respectively. (c) The distribution of properties corresponding to microstructures generated by $\bmphi_{comp}$. (d) The distribution of properties corresponding to microstructures generated by $\bmphi_{prop}$.}
\label{fig:case2_inverse}
\end{figure}

In inverse materials design, it is common to only have access to data with properties distributed within a specific area in the property space $\mathcal{M}$, while the objective is to design microstructures with properties outside of this area. 
This poses a significant challenge for any inverse design method as it necessitates models that can extrapolate to microstructures and properties that have not been seen.
To assess the generalization or extrapolation ability of the proposed method, we aim to generate microstructures with  properties in a region $M_d$,  depicted in green in Figure \ref{fig:case2_inverse}, while the model is trained with data (blue dots) that do not overlap with this property region and most of them, lie far away from it.

Figures \ref{fig:case2_inverse}b-d present the  results obtained by the two methods. In particular, Figure \ref{fig:case2_inverse}b illustrates the estimated  probability from the reference dataset that microstructures generated with each $\bmphi$-value have  properties in $M_d$.
The red  and blue dot in \ref{fig:case2_inverse}b represent the solutions $\bmphi_{prop}=(8.999, 1.999)$ and $\bmphi_{comp}=(7.425,1.966)$, respectively. It is evident that $\bmphi_{prop}$ most closely approximates the reference value  of $\bmphi_{ref}=(9., 1.905)$.
Moreover, for each of  the optimal $\bmphi$ values identified, we generated $1000$ microstructures and calculated the corresponding probability through exact SP simulation.  We found that $92.4\%$ of the microstructures corresponding to $\bmphi_{prop}$ exhibit properties within $M_d$, which is close to the  $94.2\%$ attained by the reference solution $\bmphi_{ref}$ (see Figure \ref{fig:case2_inverse}d).
However, only $69.7\%$ of the microstructures corresponding to $\bmphi_{comp}$ exhibit properties within $M_d$ (see Figure \ref{fig:case2_inverse}c). This outcome highlights the strong generalization/extrapolation ability of our PSP-GEN model.
\subsection{Case 3: Small training dataset}
\label{sec:case_3}
The forward simulation is often complex, time-consuming, and expensive for most inverse materials design tasks. To mitigate the effort and cost associated with simulating the forward PSP process, it is essential for inverse design methods to perform well even with limited training data.
Therefore, we investigate the performance of the PSP-GEN with a smaller training dataset in this section.

In particular, we constructed a reduced dataset consisting of only $N_{train}=2500$ triads $(\bmphi, \bmx, \bmm)$ in contrast to the $10,000$ used previously. For each of the $2500$ $\bmphi$-values considered, we generated only a single  microstructure and computed its properties.
We subsequently  solved the  three inverse design tasks discussed previously corresponding to the three target regions $M^L_d, M^M_d$, and $M^R_d$ of section \ref{sec:case_1}, using both the proposed and competitive methods.
The numerical results are presented in Table \ref{tab:case_3}.

\begin{figure}[t] \centering
\begin{overpic}[width=0.315\textwidth]{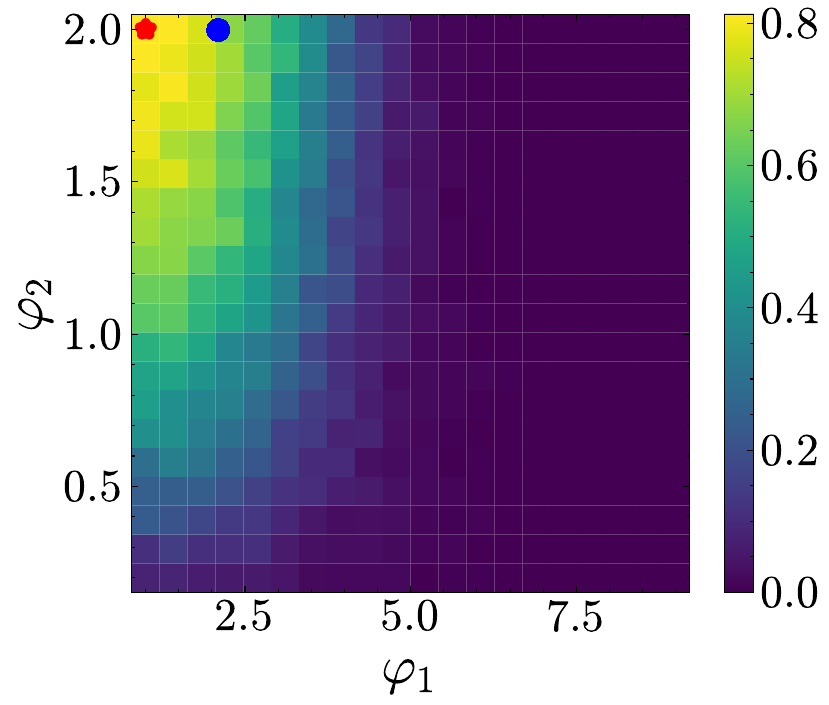}
\put(0,80){a}
\end{overpic}
\hspace{0.05cm}
\begin{overpic}[width=0.315\textwidth]{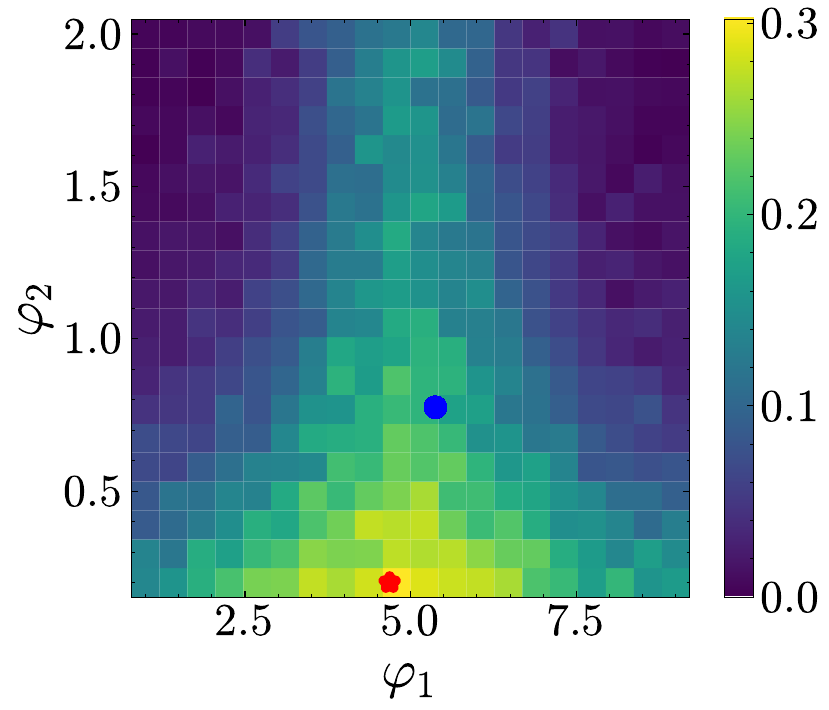}
\put(0,80){b}
\end{overpic}
\hspace{0.05cm}
\begin{overpic}[width=0.315\textwidth]{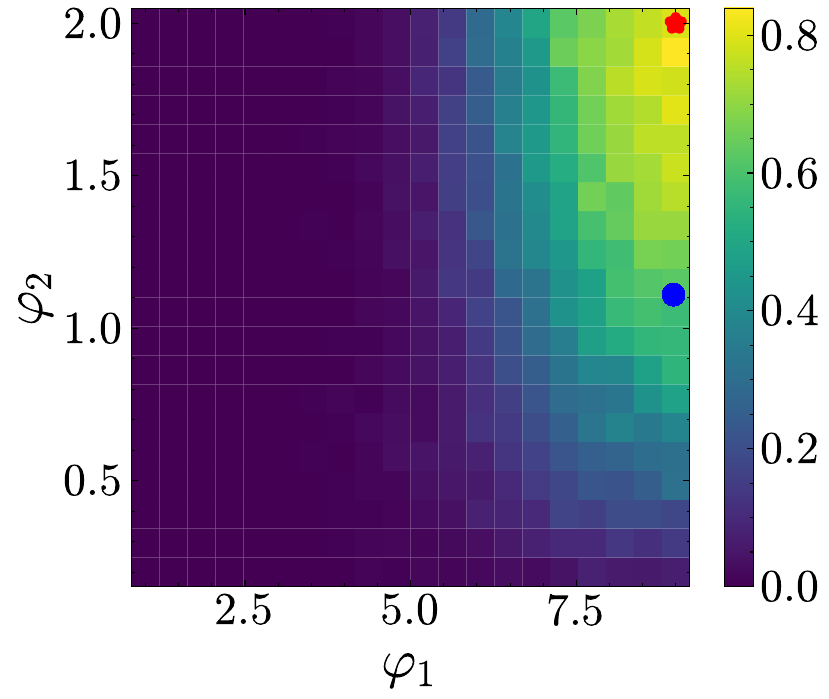}
\put(0,80){c}
\end{overpic}
\caption{The inverse design results in Case 3. (a) The green shaded areas indicate the three target regions $M^L_d$, $M^M_d$, and $M^R_d$. The dots represent property values  in the testing dataset. (b), (c), (d): From left to right, the figures correspond to the estimated  probabilities  of attaining properties in the  target regions $M^L_d$, $M^M_d$, and $M^R_d$, respectively. The estimates are computed using the reference dataset. The blue dots represent $\bmphi_{comp}$ obtained by the competitive method and the red dots the $\bmphi_{prop}$ obtained by the proposed method.}
\label{fig:case3_inverse}
\end{figure}

\begin{table}[h]
\centering
\caption{The optimal process parameters obtained by the competitive method ($\bmphi_{comp}$) and the proposed PSP-GEN ($\bmphi_{prop}$) for inverse design tasks with different target region $M_d$ in Case 3. The $\bmphi_{ref}$ denotes the reference ground truth of the processing parameter.}
\begin{tabular}{cccc} \bottomrule
$\bmphi^*$/$p_{M_d}$& $M^L_d$ &$ M^M_d$ &	$M^R_d$ \\ \bottomrule
$\bmphi_{comp}$/$p_{M_d}$& (2.095, 1.999)/$67.2\%$& (5.373, 0.775)/$19.6\%$& (8.967, 1.110)/$62.4\%$\\ 
$\bmphi_{prop}$/$p_{M_d}$& (1.001, 1.999)/$81.2\%$& (4.683, 0.200)/$30.2\%$&	(8.999, 1.999)/$79.2\%$	\\ 
$\bmphi_{ref}$/$p_{M_d}$& (1.000, 2.000)/$81.2\%$&	(4.789, 0.200)/$30.2\%$&	(9.000, 1.905)/$83.8\%$	\\ \toprule
\end{tabular}
\label{tab:case_3}
\end{table}
Additionally, we visualize the performance of the two methods in the target regions $M^L_d$, $M^M_d$, and $M^R_d$ in Figures \ref{fig:case3_inverse}a, \ref{fig:case3_inverse}b, and \ref{fig:case3_inverse}c, respectively.
As depicted in Figure \ref{fig:case3_inverse}, for all three tasks, the $\bmphi_{prop}$ obtained by the PSP-GEN model more closely approximates  the reference solution. Furthermore, in comparison to Section \ref{sec:case_1} and Figure \ref{fig:case1_inverse} we observe no degradation in the accuracy of the proposed method on the same objectives despite the reduction in training dataset size by a factor of 4.
However, for the competitive method, the performance worsens for the target regions $M^M_d$ and $M^R_d$ as the $\bmphi_{comp}$ obtained is  further from the reference value as compared to the case with larger datasets in Section \ref{sec:case_1}. Although its performance appears to improve for the target region $M^L_d$, the obtained solution is still far from the optimal solution. This indicates that the competitive method is more sensitive to the size of the training dataset than the proposed method.

\subsection{Case 4: High-dimensional microstructure space}
\label{sec:case_4}
In this section, we explore the efficiency of the proposed method in addressing the inverse design problem with high-dimensional microstructure spaces.
To this end,  we increased the image size to $128\times 128$ pixels.
We employed the same PS and SP models described in section \ref{sec:dataset} to generate the training, testing, and reference datasets (we set the parameter $\gamma$ in \refeqp{eq:CH_eq} to $\gamma=10$). 

\begin{figure}[t] \centering
\begin{overpic}[width=0.22\textwidth]{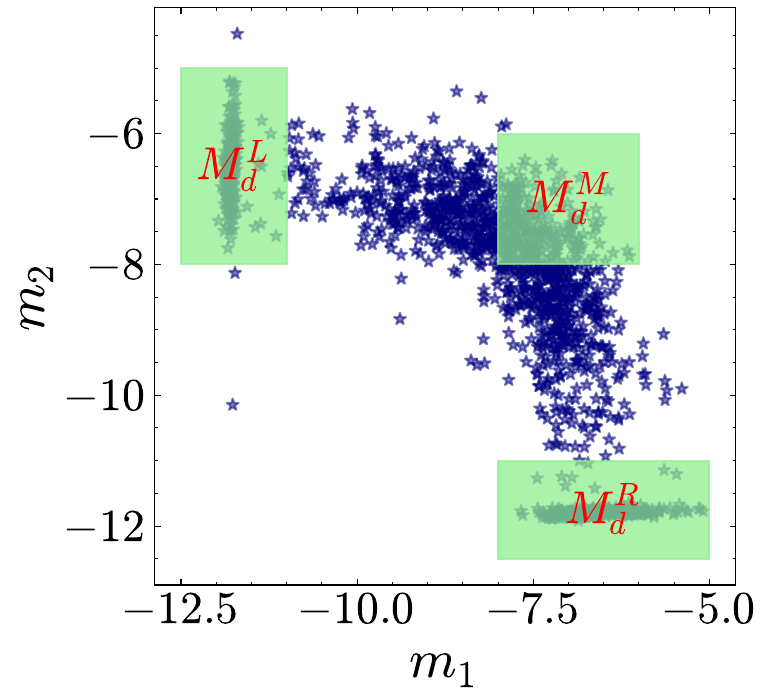}
\put(0,90){a}
\end{overpic}
\hspace{-0.05cm}
\begin{overpic}[width=0.245\textwidth]{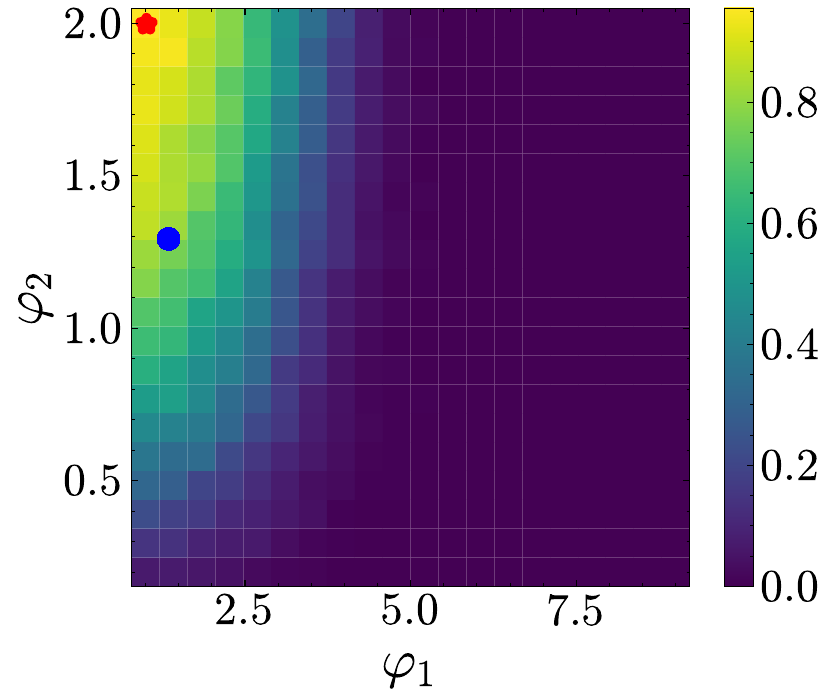}
\put(0,80){b}
\end{overpic}
\hspace{-0.05cm}
\begin{overpic}[width=0.245\textwidth]{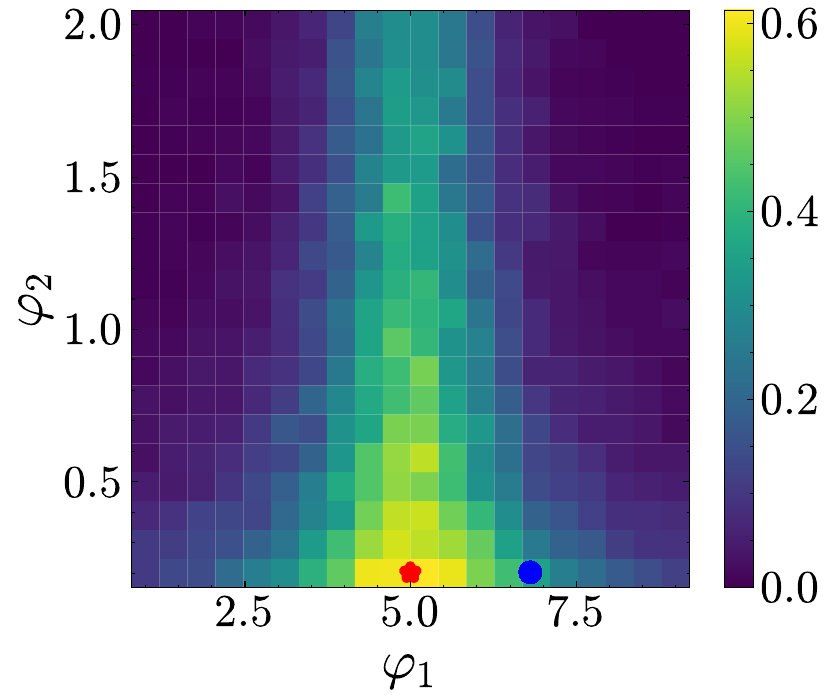}
\put(0,80){c}
\end{overpic}
\hspace{-0.05cm}
\begin{overpic}[width=0.245\textwidth]{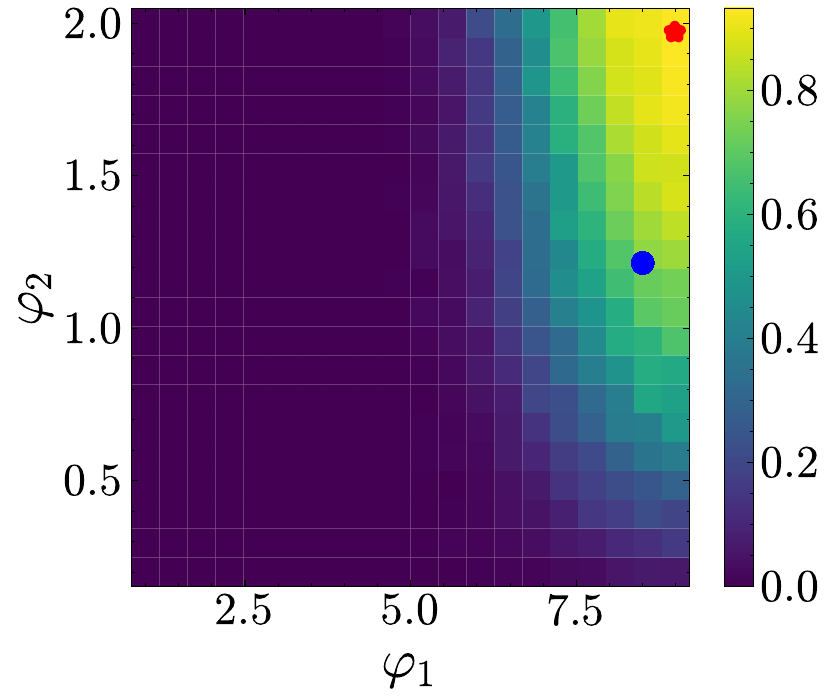}
\put(0,80){d}
\end{overpic}
\caption{The inverse design results in Case 4. (a) The green shaded areas indicate the three target regions $M^L_d$, $M^M_d$, and $M^R_d$. The dots represent property values  in the testing dataset. (b), (c), (d): From left to right, the figures correspond to the estimated  probabilities  of attaining properties in the  target regions $M^L_d$, $M^M_d$, and $M^R_d$, respectively. The estimates are computed using the reference dataset. The blue dots represent $\bmphi_{comp}$ obtained by the competitive method and the red dots the $\bmphi_{prop}$ obtained by the proposed PSP-GEN.}
\label{fig:case5_inverse}
\end{figure}

Similar to Section \ref{sec:case_1}, we considered  three,  different, target property domains, namely $M^L_d=[-12.5, -11]\times[-8,-5]$, $M^M_d=[-8,-6]\times[-8,-6]$, and $M^R_d=[-8,-5]\times[-12.5,-11]$,  as shown in Figure \ref{fig:case5_inverse}a.
We train the forward model and subsequently solve the inverse problem for all three cases. The optimal processing parameters obtained by both methods for these target regions are recorded in Table \ref{tab:case_5}.
We also visualize the results in target regions $M^L_d$, $M^M_d$, and $M^R_d$ in Figure \ref{fig:case5_inverse}b, \ref{fig:case5_inverse}c, and \ref{fig:case5_inverse}d, respectively.
As depicted in Figure \ref{fig:case5_inverse}, for all three target regions, the solution $\bmphi_{prop}$ obtained by the PSP-GEN is more accurate than the solution $\bmphi_{comp}$ obtained by the competitive method, and closely aligns with the reference values $\bmphi_{ref}$.
%

\begin{figure}[!t] 
\centering
\hspace{-0.05cm}
\begin{overpic}[width=0.675\textwidth]{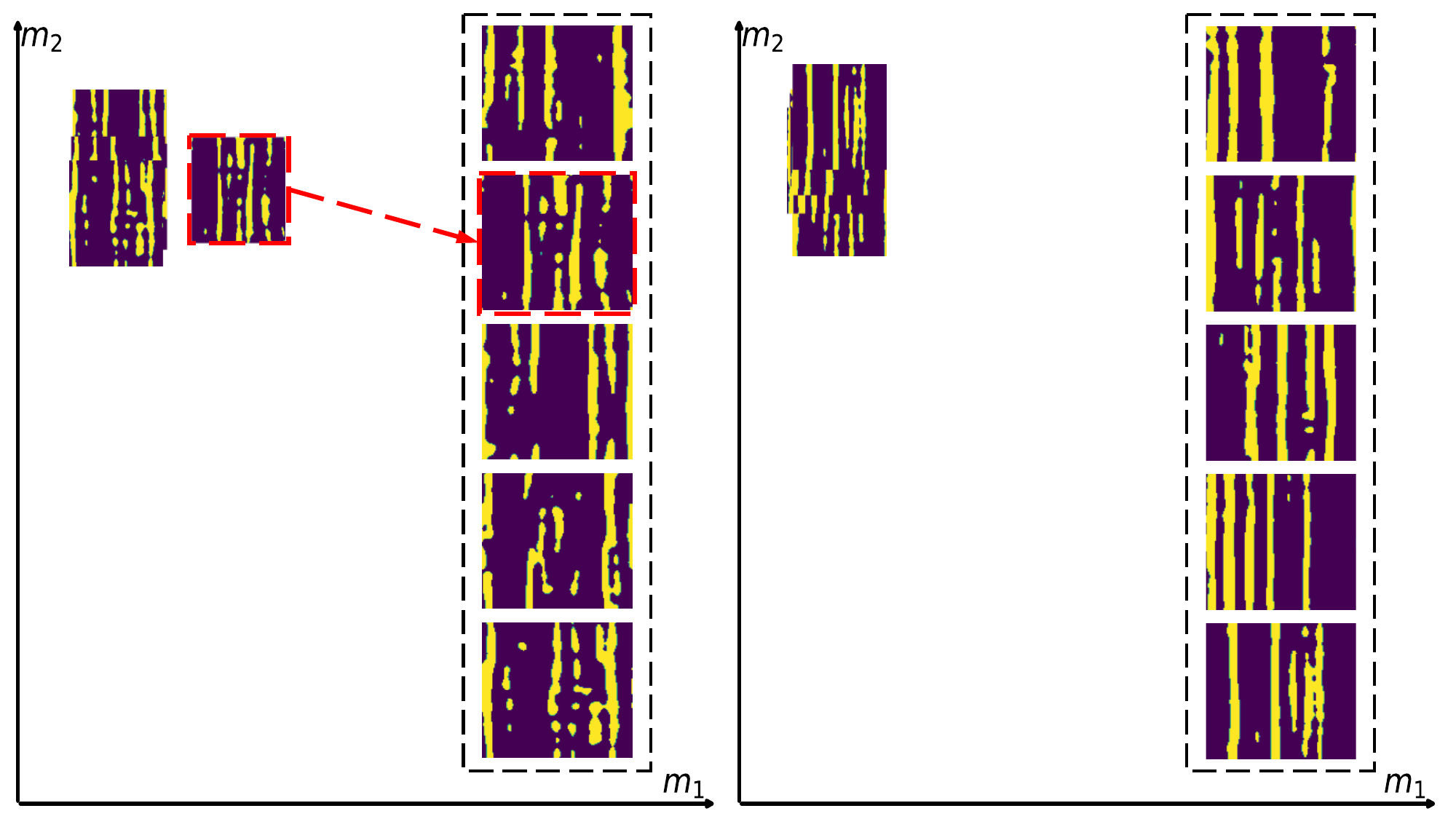}
\put(-2.5, 53){a}
\end{overpic}
\hspace{-0.05cm}
\begin{overpic}[width=0.675\textwidth]{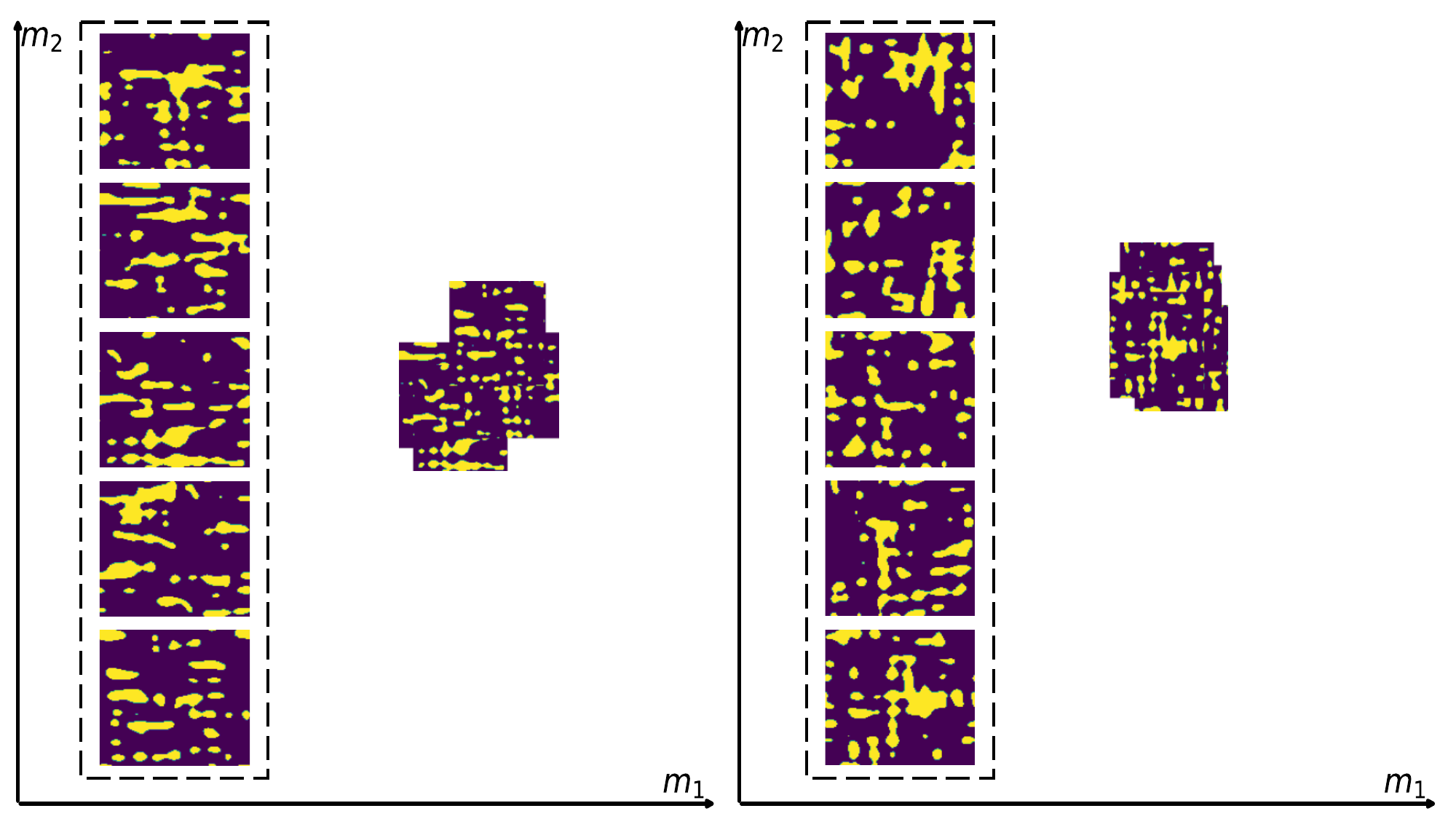}
\put(-2.5,53){b}
\end{overpic}
\hspace{-0.05cm}
\begin{overpic}[width=0.675\textwidth]{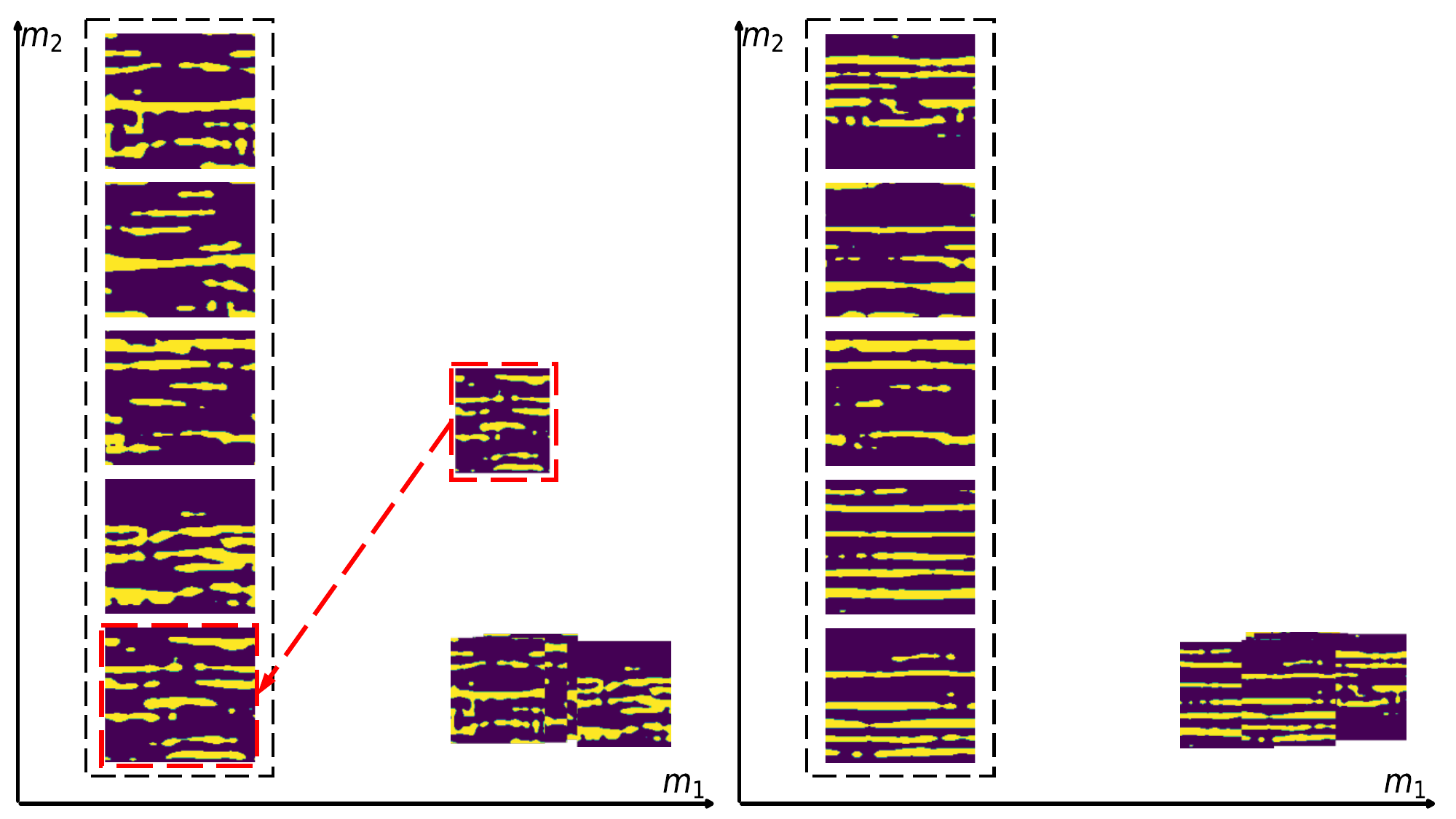}
\put(-2.5,53){c}
\end{overpic}
\caption{The microstructures generated with optimal processing parameters obtained by different methods in Case 4. (a) microstructures generated with $\bmphi_{comp}$ (left) and $\bmphi_{prop}$ (right) for target region $M^L_d$. (b) microstructures generated with $\bmphi_{comp}$ (left) and $\bmphi_{prop}$ (right) for target region $M^M_d$. (c) microstructures generated with $\bmphi_{comp}$ (left) and $\bmphi_{prop}$ (right) for target region $M^R_d$. The solid phase is indicated in yellow.}
\label{fig:case4_check_x}
\end{figure}

Finally, in Figure \ref{fig:case4_check_x}, which could be thought of as a higher resolution version of  Figure \ref{fig:case1_check_x},   we plot for each target region a few indicative microstructures generated by the   identified $\bmphi_{comp}$ (left) and $\bmphi_{prop}$ (right).
For the target region $M^L_d$, our goal is to generate microstructures with low permeability in the horizontal direction. Figure \ref{fig:case4_check_x}a shows that the microstructures generated using $\bmphi_{prop}$ all exhibit the characteristic of the solid phase (indicated with yellow) forms connected  paths along the vertical direction, resulting in low horizontal permeability  targeted. While the microstructures generated by  $\bmphi_{comp}$ also show similar characteristics, in some of them (indicated by a red dashed box) the solid phase's paths exhibit intermittency, leading to  connected void paths (denoted with black), and thus resulting in relatively higher permeability along the horizontal direction.
Similarly, for the target region $M^R_d$, we aim to generate microstructures with low permeability in the vertical direction, meaning the solid phase (denoted with yellow) should be distributed so as to block as much as possible flow in the vertical direction. In Figure \ref{fig:case4_check_x}c, we  observe that all microstructures generated by $\bmphi_{prop}$ exhibit this characteristic. In contrast, some of  the microstructures generated by $\bmphi_{comp}$ (shown  within a red dashed box) do not exhibit the desired characteristic resulting in higher permeability in the vertical direction than targeted. 
For the target region $M^M_d$, the goal is to generate microstructures with high permeability in both directions which should favor some connected paths of the void phase (indicated with black) along both directions. As seen in Figure \ref{fig:case4_check_x}b, the microstructures generated by $\bmphi_{prop}$ all exhibit this characteristic. In contrast, the microstructures generated by $\bmphi_{comp}$ show a tendency, albeit weak, for the solid phase to be distributed horizontally. This trend can be observed in the distribution of microstructures in the $m_1-m_2$ coordinate, where the microstructures generated by $\bmphi_{comp}$ are more dispersed and tend towards the lower permeability region compared to $\bmphi_{prop}$.
\begin{table}
\centering
\caption{The optimal process parameters obtained by the competitive method ($\bmphi_{comp}$) and the proposed PSP-GEN ($\bmphi_{prop}$) for inverse design tasks with different target region $M_d$ in Case 4. The $\bmphi_{ref}$ denotes the reference ground truth of the processing parameter.}
\begin{tabular}{cccc} \bottomrule
$\bmphi^*$/$p_{M_d}$& $M^L_d$ &$ M^M_d$ &	$M^R_d$ \\ \bottomrule
$\bmphi_{comp}$/$p_{M_d}$& (1.347, 1.292)/$81.6\%$& (6.803, 0.203)/$35.2\%$& (8.502, 1.213)/$77.6\%$\\
$\bmphi_{prop}$/$p_{M_d}$& (1.013, 1.997)/$95.4\%$& (4.996, 0.202)/$60.2\%$&	(8.987, 1.971)/$93.2\%$	\\
$\bmphi_{ref}$/$p_{M_d}$& (1.000, 2.000)/$95.4\%$&	(5.211, 0.200)/$61.4\%$&	(9.000, 2.000)/$93.2\%$	\\ \toprule
\end{tabular}
\label{tab:case_5}
\end{table}

\subsection{Case 5: High-dimensional processing parameter space}
\label{sec:case_5}
\begin{figure}[!t] \centering
\begin{overpic}[width=1.0\textwidth]{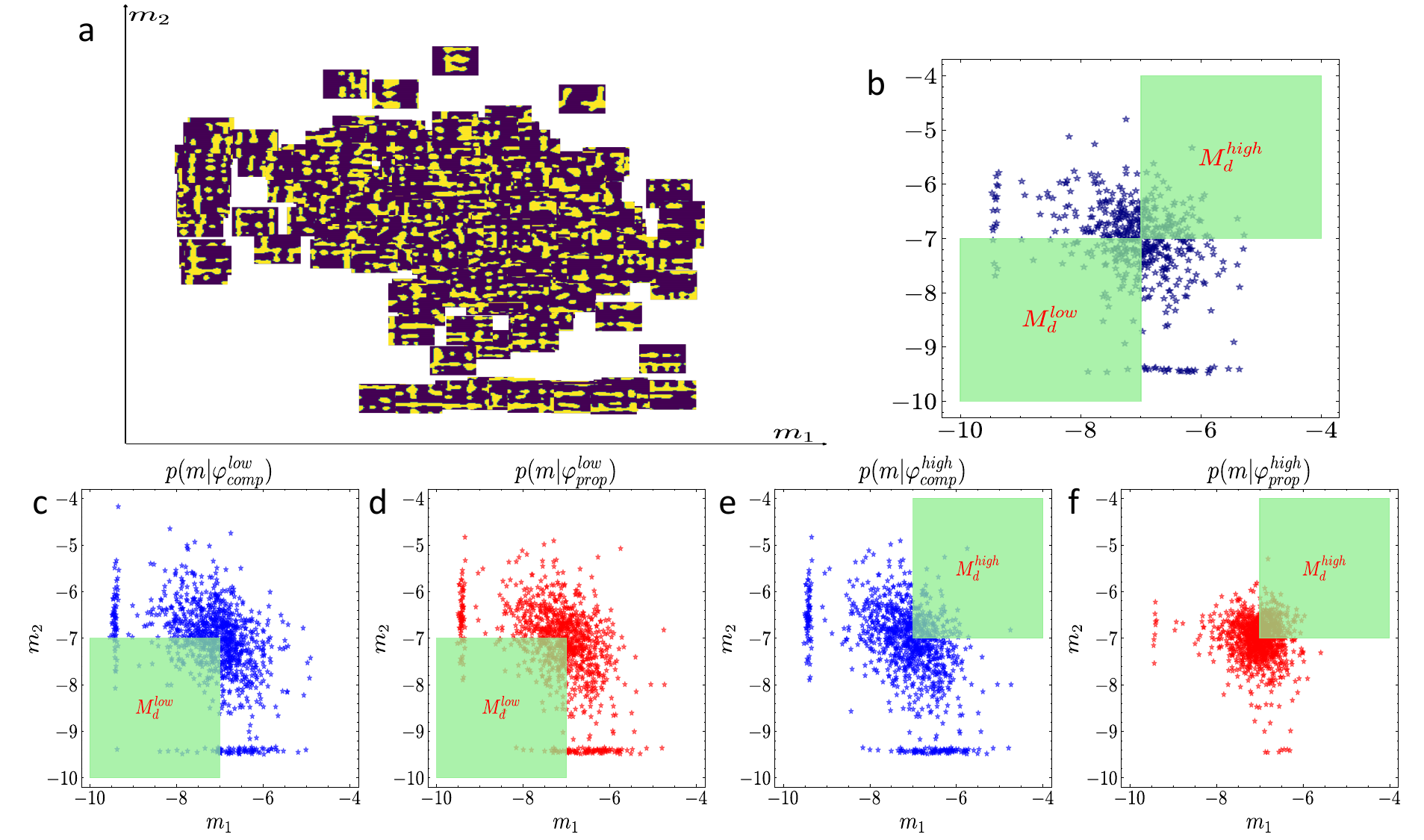}
\end{overpic}
\caption{The inverse design results in Case 5. (a) The distribution of microstructures $\bmx$ in the testing dataset, with locations determined by the corresponding properties. (b) The properties $\bmm$ corresponding to the microstructures $\bmx$. (c) The distribution of properties corresponding to microstructures generated by $\bmphi^{low}_{comp}$. (d) The distribution of properties corresponding to microstructures generated by $\bmphi^{low}_{prop}$. (e) The distribution of properties corresponding to microstructures generated by $\bmphi^{high}_{comp}$. (f) The distribution of properties corresponding to microstructures generated by $\bmphi^{high}_{prop}$.}
\label{fig:case4_task}
\end{figure}

In this section, we evaluate the performance of the proposed method in addressing inverse materials design problems with a high-dimensional process parameter space.
This is an inherently challenging task as it entails navigating a high-dimensional design space to identify optimal solutions.
To generate data for this high-dimensional problem, we adapt the SDF in the GRF as follows
\begin{equation}
 S(\bmphi) = \sum^{5}_{i=1}(\varphi_{i,1} * e^{-\varphi_{i,2} * k_x^2} + \varphi_{i,1} *  e^{-\varphi_{i,2} * k_y^2}),\quad \bmphi\in\mathcal{U}_{\varphi}=[1, 9]^{5}\times[0.02, 2]^{5}
\end{equation}
where $\bmphi=(\varphi_{1,1}, \cdots, \varphi_{5,1}, \varphi_{1,2}, \cdots, \varphi_{5,2})$ represents a ten-variable vector. The remaining settings for the PSP process remain the same as in section \ref{sec:case_1}. We sample $10,000$ process parameter values from $\mathcal{U}_{\varphi}$ and generate one microstructure for each parameter.
Subsequently, we compute the properties of these microstructures, yielding $N=10,000$ pairs of $(\bmphi, \bmx, \bmm)$ for training purposes.
For the testing dataset, we sample $500$ process parameters from $\mathcal{U}_\varphi$ and generate $4$ microstructures for each $\bmphi$. We then compute the properties for these microstructures to get a testing dataset with size $N=2,000$.
Figures \ref{fig:case4_task}a and \ref{fig:case4_task}b show the microstructures and corresponding properties in the testing dataset, respectively. Compared with the low-dimensional processing parameter space case, we observed that the microstructures are more concentrated. This is because the expanded dimensions of the processing parameters make the microstructural space corresponding to moderate permeability in both horizontal and vertical directions larger.
Owing to the high dimensionality of the processing parameter space, it becomes impractical to generate a large reference dataset to obtain a reference solution of the inverse design
problem.
Therefore, we assess the performance of the methods through the following task: We define a target region $M_d$ and solve the inverse problem to obtain the optimal process parameter $\bmphi^*$. We then utilize the $\bmphi^*$ to generate microstructures via real PSP simulation and compute the percentage of microstructures with properties in $M_d$, i.e. $p_{M_d}:=p(\bmm\in M_d|\bmphi^*)$.

To make the task challenging, we considered two complex target regions (see Figure \ref{fig:case4_task}b): the target region $M_d^{low} = [-10, -10] \times [-7, -7]$, which  corresponds to microstructures with low permeability in both the horizontal and vertical directions; and the target region $M_d^{high} = [-7, -7] \times [-4, -4]$, which  corresponds to microstructures with high permeability in both directions. The solutions obtained by our method and the competitive method for $M_d^{low}$ are as follows:
\be
\begin{array}{l}
\bmphi^{low}_{prop}=(1.000, 1.000, 1.000, 3.092, 1.000, 0.020, 0.020, 0.020, 0.020, 2.000), \nonumber \\
\bmphi^{low}_{comp}=(1.046, 8.995 4.802, 9.000, 5.036, 1.999, 2.000, 2.000, 2.000, 1.936).
\end{array}
\nonumber
\ee
For $M_d^{high}$, the solutions obtained by the two methods are as follows:
\be
\begin{array}{l}
\bmphi^{high}_{prop}=(1.000, 3.055, 2.374, 8.952, 8.999, 0.020, 0.020, 0.024, 0.021, 0.095), \\
 \bmphi^{high}_{comp}=(6.183, 7.512, 4.351, 8.999, 8.999, 2.000, 0.022, 1.999, 0.023, 2.000).
 \end{array}
\ee
To evaluate their performance, we use the obtained solutions to generate $1000$ microstructures and calculate the percentage of microstructures with properties in the target region, as shown in Figure \ref{fig:case4_task}c-f.
For the target region $M_d^{low}$, the results are $p_{M_d^{low}}=19.2\%$ and $p_{M_d^{low}}=18\%$ for the proposed PSP-GEN method and the competitive method, respectively. For the target region $M_d^{high}$, the results are $p_{M_d^{high}}=24.7\%$ and $p_{M_d^{high}}=14.2\%$ for the proposed PSP-GEN method and the competitive method, respectively. On both tasks and for the 10-dimensional processing design space, we observe the superior performance of the PSP-GEN method to the competitive one.
\section{Concluding remarks}
\label{sec:conclusion}
In summary, we have introduced an advanced framework, PSP-GEN, for goal-oriented materials design that comprehensively models the entire PSP chain using a deep, generative model. The PSP-GEN addresses numerous challenges inherent in inverse materials design, including the presence of stochasticity along the whole modeling chain, the high-dimensional, discrete-valued microstructure space that precludes the availability of derivatives, and the complex and nonlinear SP process.

Central to the effectiveness of the PSP-GEN framework is the continuous,  latent space which consists of two components. The first provides a lower-dimensional representation that captures the stochastic aspects of microstructure generation, while the second offers a direct link to processing parameters. This structured, low-dimensional embedding not only simplifies the handling of high-dimensional microstructure data but also facilitates the application of gradient-based optimization techniques. Moreover, the incorporation  of the microstructures and properties  in learning the  latent generators ensures that our method is microstructure- and property-aware, leading to more precise solutions compared to microstructure-agnostic methods or those that employ dimensionality reductions solely in the microstructural space.

The PSP-GEN framework significantly enhances computational efficiency and precision in inverse material design. By transforming the goal-oriented design problem into a stochastic,  optimization problem with continuous optimization variables, we achieve superior accuracy compared to the recent state-of-the-art method, particularly in settings where one must operate with limited training data or with high-dimensional parameter spaces. Furthermore, our framework demonstrates strong generalization ability in cross-region design problems, i.e. where the target, property  region is far away from the training data space.

Beyond the specific case studies presented, Our PSP-GEN framework holds promise for widespread application in materials science. Its ability to integrate detailed microstructural information and effectively manage stochastic and deterministic elements positions it as a potent tool for accelerating material discovery and optimization across industries. One acknowledged limitation is the substantial training data typically required for the generative model. Future research efforts should focus on enhancing the efficiency of the PSP-GEN framework by incorporating physical information directly into the latent representations, thereby reducing the data requirements while maintaining or improving performance. This advancement would further solidify the framework’s utility and broaden its impact in advancing materials science research and development.



\newpage
\bibliographystyle{elsarticle-num}
\bibliography{ref.bib}


\appendix
\section{The model structures}
\label{apdx:model_structure}
\paragraph{The model $p_{\bmtheta_\bmzb}(\bmzb|\bmphi)$} For the model $p_{\bmtheta_\bmphi}(\bmzb|\bmphi)=\delta(\bmzb - g_{\bmtheta_\bmphi}(\bmphi))$, we modeled $g_{\bmtheta_\bmphi}(\bmphi)$ as a ResNet model consists of an input layer, three hidden layers with 256, 128, and 64 neurons respectively using ELU activation, and an output layer with linear activation.
\paragraph{The Decoder $D_{\bmtheta_\bmx}$} For the decoder $\bm{\mu}_{\bmtheta_\bmx}=D_{\bmtheta_\bmx}(\bmza, \bmzb)$, we modeled it as a ResNet model with an input layer, two hidden layers with 512 and 128 neurons respectively using ReLU activation, and an output layer.
At the output layer, the sigmoid activation was applied to transform the output value into $(0,1)$ for probabilistic outputs.
\paragraph{The Decoder $D_{\bmtheta_\bmm}$} For the decoder $(\bm{\mu}_{\bmtheta_\bmm},\bm{\sigma}_{\bmtheta_\bmm})= D_{\bmtheta_\bmm}(\bmza, \bmzb)$, we selected both models as a ResNet with an input layer, three hidden layers with 64, 128 and 256 neurons respectively using ELU activation, and an output layer with linear activation.
\paragraph{The Encoder $E_{\bmeta}$:} For the encoder $\bmza = E_{\bmtheta_\bmm}(\bmm, \bmx, \bmphi)$, since the input terms $\bmm$, $\bmx$, and $\bmphi$ have different data formats and scales, we first map them into three tensors of dimension $d_{\bmza}$ using three different ResNet models: \textit{ResNet}$_\bmx$, \textit{ResNet}$_\bmm$, and \textit{ResNet}$_\bmphi$.
The \textit{ResNet}$_\bmx$ is a ResNet model consisting of an input layer, two hidden layers, and an output layer. The two hidden layers have 128 and 512 neurons, respectively, and use the ELU activation function, while the output layer uses linear activation.
The \textit{ResNet}$_\bmm$ and \textit{ResNet}$_\bmphi$ have the same structure, which consists of a ResNet model with an input layer, one hidden layer, and an output layer. The hidden layer has 64 neurons and uses the ELU activation function, while the output layer uses linear activation.
Then, we concatenate these three tensors and pass them to a ResNet model with an input layer, one hidden layer, and an output layer. The hidden layer has 64 neurons and uses the ELU activation function, while the output layer uses linear activation.

Although we used the above network structure in this paper, it is important to note that the chosen network structure is not necessarily the optimal one. The optimal network structure depends on various factors, such as the representation of the microstructure, the dimensions of the microstructure space, and the objectives of the inverse problem. However, this is beyond the scope of this paper and will be explored in future research.
\section{Trade-off between reconstruction terms and the KL-divergence term}
\label{apdx:beta}
\setcounter{figure}{0}
\begin{figure}[!t] \centering
\begin{overpic}[width=0.6\textwidth]{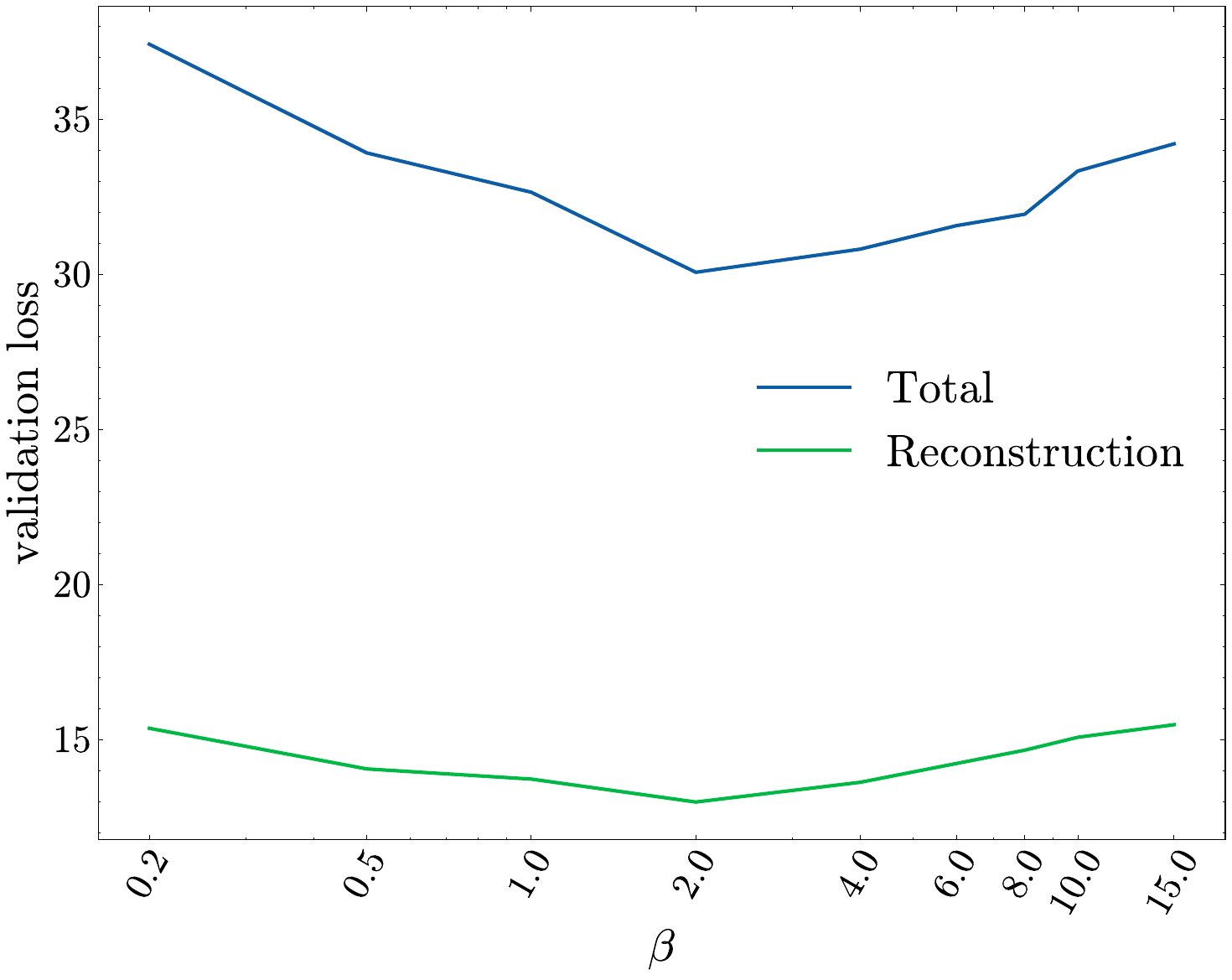}
\end{overpic}
\caption{Trade-off between reconstruction terms and the KL-divergence term: The blue curve represents the total validation losses $\mathcal{L}$ obtained by training the model with different values of $\beta$. The green curve shows the reconstruction validation losses, specifically $\mathcal{L}_{\textrm{REC}, \bmx} + \mathcal{L}_{\textrm{REC}, \bmm}$, under varying values of $\beta$.}
\label{fig:case1_beta}
\end{figure}
To obtain disentangled latent representations, we balance the terms in the ELBO using the $\beta$-VAE trick. Specifically, we selected different $\beta$ values in the objective function shown in equation \refeqp{eq:elboibeta} and trained the PSP-GEN model. For each $\beta$ value, we recorded the total validation loss $\mathcal{L}=\mathcal{L}_{REC,\bm{x}}+\mathcal{L}_{REC,\bm{m}}+\mathcal{L}_{KL}$ and the reconstruction validation loss $\mathcal{L}_{REC}=\mathcal{L}_{REC,\bm{x}}+\mathcal{L}_{REC,\bm{m}}$ obtained after the training, and plotted their curves with respect to $\beta$ in Figure \ref{fig:case1_beta}. From the figure, we can see that both the total validation loss and the reconstruction validation loss reach their minimum values when $\beta=2$. Therefore, we used the setting of $\beta=2$ in our experiments.


\end{document}